\documentclass[aps,prd,preprint,tightenlines,superscriptaddress,
   showpacs,nofootinbib]{revtex4}
\newcommand{\PRE}[1]{{#1}} 

\usepackage{bm} \usepackage{epsfig}

\usepackage{slashed}

\newcommand{\slam}{\slashed}
\newcommand{\be}{\begin{equation}}
\newcommand{\ee}{\end{equation}}
\newcommand{\beq}{\begin{equation}}
\newcommand{\eeq}{\end{equation}}
\newcommand{\beqa}{\begin{eqnarray}}
\newcommand{\eeqa}{\end{eqnarray}}

\newcommand{\eqref}[1]{Eq.~(\ref{#1})}

\newcommand{\secref}[1]{Sec.~\ref{sec:#1}}

\newcommand{\appref}[1]{Appendix~\ref{app:#1}} 

\newcommand{\Figref}[1]{Figure~\ref{fig:#1}}
\newcommand{\figref}[1]{Fig.~\ref{fig:#1}}
\newcommand{\figsref}[2]{Figs.~\ref{fig:#1} and \ref{fig:#2}}

\newcommand{\intps}{\int_{\text{PS}_3}}

\def\slashchar#1{\setbox0=\hbox{$#1$}           
   \dimen0=\wd0                                 
   \setbox1=\hbox{/} \dimen1=\wd1               
   \ifdim\dimen0>\dimen1                        
      \rlap{\hbox to \dimen0{\hfil/\hfil}}#1
   \else                                        
      \rlap{\hbox to \dimen1{\hfil$#1$\hfil}}/
       \fi}
\def\notslashchar#1{\setbox0=\hbox{$#1$}           
   \dimen0=\wd0                                 
   \setbox1=\hbox{/} \dimen1=\wd1               
   \ifdim\dimen0>\dimen1                        
      \rlap{\hbox to \dimen0{\hfil\phantom{/}\hfil}}#1
   \else                                        
      \rlap{\hbox to \dimen1{\hfil$#1$\hfil}}/
       \fi}

\begin{document}

\preprint{UCI-TR-2009-07}

\title{
\PRE{\vspace*{1.5in}}
Three-Body Decays of Sleptons with General Flavor Violation and
Left-Right Mixing
\PRE{\vspace*{0.3in}}
}

\author{Jonathan L.~Feng}
\affiliation{Department of Physics and Astronomy, University of
California, Irvine, CA 92697, USA
\PRE{\vspace*{.2in}}
}

\author{Iftah Galon}
\affiliation{Physics Department, Technion-Israel Institute of
Technology, Haifa 32000, Israel
\PRE{\vspace*{.5in}}
}

\author{David Sanford}
\affiliation{Department of Physics and Astronomy, University of
California, Irvine, CA 92697, USA
\PRE{\vspace*{.2in}}
}

\author{Yael Shadmi}
\affiliation{Physics Department, Technion-Israel Institute of
Technology, Haifa 32000, Israel
\PRE{\vspace*{.5in}}
}

\author{Felix Yu\PRE{\vspace*{.5in}}}
\affiliation{Department of Physics and Astronomy, University of
California, Irvine, CA 92697, USA
\PRE{\vspace*{.2in}}
}

\date{April 2009}

\begin{abstract}
\PRE{\vspace*{.3in}} We determine the widths of three-body decays of
sleptons, $\tilde{\ell}^- \to \tilde{\ell}^{\pm} \ell^- \ell^{\mp},
\tilde{\ell}^- \nu \bar{\nu}, \tilde{\ell}^- q \bar{q}$, in the
presence of arbitrary slepton flavor violation and left-right
mixing. These decays are important in scenarios in which the lightest
supersymmetric particle is the gravitino, a generic possibility in
models with gauge- and gravity-mediated supersymmetry breaking.
Three-body decays have been discussed previously assuming flavor
conservation and left-right mixing in only the stau sector.  Flavor
violation and general left-right mixing open up many new decay
channels, which provide new avenues for precision mass measurements
and may play an essential role in solving the standard model flavor
problem.  We present results for toy models with two-generation
mixing, and discuss the implementation of these results in {\tt
SPICE}, a program that simplifies collider event simulations of
flavor-violating supersymmetric models.
\end{abstract}


\maketitle


\section{Introduction}
\label{sec:introduction}

Fermion masses are one of the least understood parts of the standard
model (SM). Even the charged fermion masses span over five orders of
magnitude from the top quark to the electron, begging for a
theoretical explanation.  The majority of these masses and mixing
angles are precisely constrained by experiment. Still, this wealth of
data does not conclusively single out any theory of flavor.

In the near future, this may change dramatically with the discovery of
new particles at the weak scale.  New particles may only deepen the
mystery, as would be the case if a fourth generation were discovered.  On
the other hand, the masses and mixings of the new particles may be
governed by the same principles that determine the SM fermion masses.
In this case, rather than extending the fermion sector, the new
particles will shed light on the existing fermion spectrum.

Weak-scale supersymmetry provides examples of both possibilities.  In
pure gauge-mediated models, for example, squark and slepton masses are
set by flavor-blind contributions, with no connection to the SM
fermion masses.  However, in gravity-mediated models and hybrid models
with both gauge- and gravity-mediated contributions, squark and
slepton masses may receive contributions that are governed by flavor
symmetries that also determine the SM fermion
masses~\cite{Froggatt:1978nt,Nir:1993mx,Grossman:1995hk}.  The latter
possibility leads to non-trivial flavor effects in high-energy
experiments~\cite{ArkaniHamed:1996au,ArkaniHamed:1997km,Agashe:1999bm},
opening the possibility for real progress on the SM flavor problem at
the Large Hadron Collider, as has been emphasized recently by many
authors (see, for example, Refs.~\cite{BarShalom:2007pw,%
Feng:2007ke,Kribs:2007ac,Nomura:2007ap,Nomura:2008pt,BarShalom:2008fq,%
Nomura:2008gg,Kaneko:2008re,Hiller:2008sv,Hisano:2008ng}).

Here we study the implications of flavor violation for the three-body
decays of charged sleptons.  Such processes are most relevant for
colliders in models with a gravitino lightest supersymmetric particle
(LSP) and a slepton next-to-lightest supersymmetric particle (NLSP), a
generic possibility in models with both gauge-mediated supersymmetry
breaking~\cite{Dimopoulos:1996vz,Feng:1997zr} and gravity-mediated
supersymmetry breaking~\cite{Feng:2003xh,%
Feng:2003uy,Ellis:2003dn,Feng:2004zu,Feng:2004mt}.  We consider a
situation where the lightest particles are a gravitino LSP, several
light sleptons, and the lightest neutralino, with the masses of each
light slepton less than the mass of the lightest neutralino.  This
results in the typical two-body decays of these light sleptons being
highly suppressed or kinematically inaccessible, leaving the
three-body decays as the dominant decay modes.  In these cases, the
three-body decays are also often the last visible step in cascade
decays of squarks and gluinos, and so they impact nearly all
supersymmetry searches and studies.

Three-body slepton decays have been studied previously in an
important, flavor-conserving case, where the authors considered
$\tilde{e}_R \to e \tau \tilde{\tau}_1$ and $\tilde{\mu}_R \to \mu
\tau \tilde{\tau}_1$, with the $\tilde{\tau}_1$ a mixture of left- and
right-handed staus~\cite{Ambrosanio:1997bq}.  These decays are
characterised by two distinct channels: a ``charge-preserving''
channel $\tilde{\ell}^- \to \tilde{\ell}^- \ell^- \ell^+$ with
opposite-sign leptons and a ``charge-flipping'' channel
$\tilde{\ell}^- \to \tilde{\ell}^+ \ell^- \ell^-$ with same-sign
leptons.  Flavor-conserving three-body decays of squarks have also
been considered~\cite{Djouadi:2000aq}, as have flavor-conserving
three-body decays with sneutrinos as parent or daughter
particles~\cite{Kraml:2007sx}.  Our work generalizes the charged
slepton analysis to the case of arbitrary lepton flavor violation
(LFV) and arbitrary left-right mixing.  In the presence of general
LFV, any three-body decay of $\tilde{\ell}_i \to \tilde{\ell}_j$ has
up to 9 possible charge-preserving modes $\tilde{\ell}_i^- \to
\tilde{\ell}_j^- \ell_k^- \ell_m^+$ and 6 possible charge-flipping
modes $\tilde{\ell}_i^- \to \tilde{\ell}_j^+ \ell_k^- \ell_m^-$, where
$\ell_k, \ell_m = e, \mu, \tau$.  In addition, LFV and left-right
mixing bring additional complications that are absent in the
flavor-conserving case, including new processes mediated by Higgs and
$Z$ bosons, new final states with neutrinos and quarks, and new
interference effects in charge-flipping processes.

These complications are well worth confronting, however, as there is a
wealth of information in these branching ratios, which may shed light
on the SM flavor problem.  These decays, if present, are also relevant
more broadly, for example, for supersymmetric searches and precision
mass measurements.  For example, in the flavor-conserving examples
studied previously~\cite{Ambrosanio:1997bq}, the final states
necessarily contain $\tau$ leptons. Since these decay with missing
energy, they degrade searches based on energetic leptons, and they
greatly reduce the prospects for precision mass measurements.  With
LFV, however, even if the lightest slepton is a stau, there may also
be decay modes with two electrons, two muons, or an electron and a muon.
Even if these branching ratios are suppressed, they may be the more
obvious signals at colliders, and they may also provide better
opportunities for precision mass measurements.  It is therefore of
interest to know the size of these branching ratios, and what
determines them.

In the following three sections, we begin with a general discussion of
three-body decays and move gradually to more specific scenarios and
concrete calculations.  In \secref{diagrams} we present the new final
states and new Feynman diagrams that are relevant to three-body decays
once general flavor and left-right mixing are introduced.  In
\secref{widths} we discuss these results in more detail and explain
the relative phenomenological importance of the various contributing
diagrams in particular scenarios.  In \secref{plots} we then show
concrete results in two toy models with two-generation slepton mixing
to illustrate our results.  Finally, in \secref{conclusion}, we
present our conclusions and explain how our results have been
incorporated into {\tt SPICE}, a publicly available computation
package for generating supersymmetric spectra and branching ratios in
scenarios with arbitrary slepton mixing.

We stress that, although we strive to give readers a intuitive feel
for our results by considering concrete cases in the body of the
paper, our analysis is valid for fully general LFV and left-right
mixing.  The complete, model-independent calculation is lengthy, but
the full expressions for all three-body decay modes are given in a
series of appendices.  Our conventions and notations are defined in
\appref{lagrangian}.  These are consistent with those of
Ref.~\cite{Engelhard:SPICE}, where full details may be found.


\section{Flavor Violation in Three-Body Slepton Decays}
\label{sec:diagrams}

\begin{figure}[tb]
\includegraphics[width=13cm]{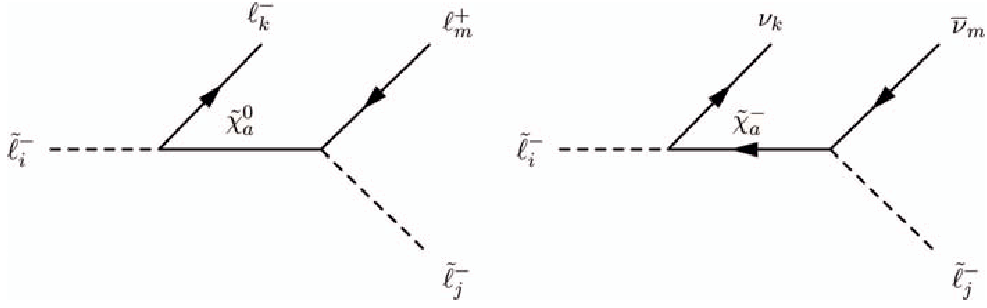}
\caption{Feynman diagrams for charge-preserving decays
$\tilde{\ell}_i^- \to \tilde{\ell}_j^- \ell_k^- \ell_m^+$ mediated by
neutralinos and $\tilde{\ell}_i^- \to \tilde{\ell}_j^- \nu_k
\overline{\nu}_m$ mediated by charginos.}
\label{fig:preserving}
\end{figure}

\begin{figure}[tb]
\includegraphics[width=13cm]{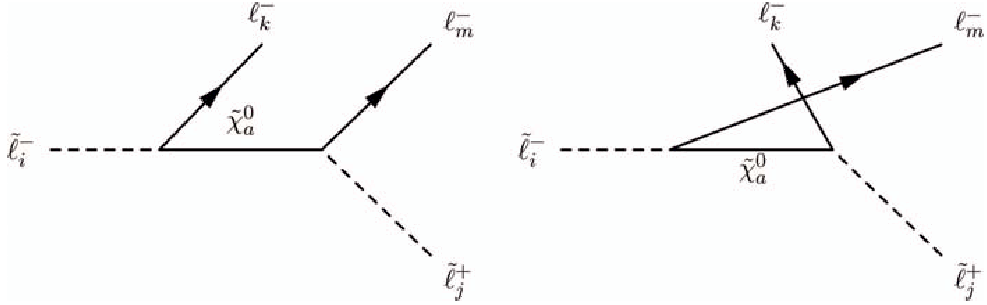}
\caption{Feynman diagrams for the charge-flipping decays
$\tilde{\ell}_i^- \to \tilde{\ell}_j^+ \ell_k^- \ell_m^-$ mediated by
neutralinos.}
\label{fig:flipping}
\end{figure}

As mentioned in the Introduction, an important special case of
three-body slepton decays has been discussed previously by Ambrosanio,
Kribs, and Martin~\cite{Ambrosanio:1997bq}.  Motivated by pure
gauge-mediated models, these authors considered the flavor-conserving
decays $\tilde{e}_R \to e \tau \tilde{\tau}_1$ and $\tilde{\mu}_R \to
\mu \tau \tilde{\tau}_1$.  These decays are mediated solely by
neutralinos.  The charge-preserving modes $\tilde{\ell}_i^- \to
\tilde{\ell}_j^- \ell_k^- \ell_m^+$ are shown in \figref{preserving},
and the charge-flipping modes $\tilde{\ell}_i^- \to \tilde{\ell}_j^+
\ell_k^- \ell_m^-$ are shown in \figref{flipping}. The charge-flipping
mode is made possible by the Majorana nature of the neutralino, and
detection of the charge-flipping mode would provide strong evidence
that neutralinos are Majorana fermions.

In the presence of more general slepton mixing, the gauge eigenstates
$\tilde{e}_R$, $\tilde{\mu}_R$, $\tilde{\tau}_R$, $\tilde{e}_L$,
$\tilde{\mu}_L$, and $\tilde{\tau}_L$ mix to form six mass
eigenstates, $\tilde{\ell}_i$, $i=1, \ldots, 6$, with increasing mass,
and the lepton-slepton-neutralino interactions are no longer
flavor-diagonal.  The neutralino diagrams of
\figsref{preserving}{flipping} are then modified by the inclusion of
$6\times 6$ mixing matrix factors at the interaction vertices.

\begin{figure}[tb]
\includegraphics[width=13cm]{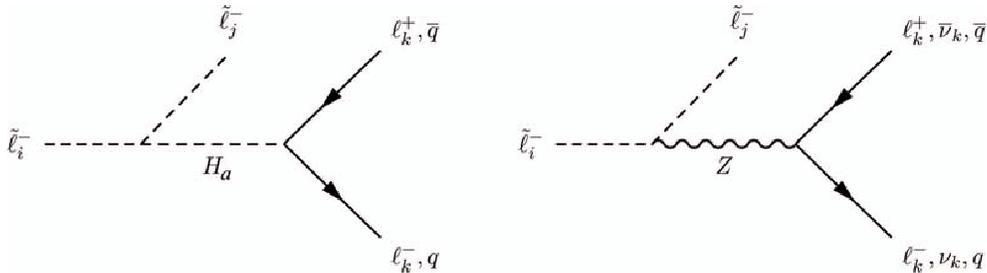}
\caption{Feynman diagrams for charge-preserving decays mediated by
Higgs and $Z$ bosons.  The Higgs scalar $H_a$ may be any of the
neutral Higgs bosons: $h^0$, $H^0$, $A^0$, or (in Feynman gauge)
$G^0$.}
\label{fig:preservingHZ}
\end{figure}

In addition, new diagrams contribute.  If the initial and final state
charged sleptons contain left-handed components, there is the
chargino-mediated decay to neutrinos shown in \figref{preserving}.
This decay requires neither flavor violation nor left-right mixing.
There are also charge-preserving decays mediated by neutral Higgs
bosons and the $Z$ boson, as shown in \figref{preservingHZ}.  The
Higgs bosons mediate decays to same flavor $\ell^+ \ell^-$ and $q
\bar{q}$ pairs, and the $Z$ diagram mediates decays to same flavor
$\nu \bar{\nu}$, $\ell^+ \ell^-$, and $q \bar{q}$ pairs.

Both the $H_a$ and $Z$ diagrams are present independent of LFV, but
both require left-right mixing.  In the case of the Higgs-mediated
diagrams, the Higgs couplings to sleptons have the form $H_a
\tilde{\ell}_L^* \tilde{\ell}_R$, $H_a \tilde{\ell}_L^*
\tilde{\ell}_L$, and $H_a \tilde{\ell}_R^* \tilde{\ell}_R$.  The first
is generated by left-right mixing terms in the Lagrangian, and thus
clearly requires left-right mixing to be non-zero.  The other two come
from $D$ terms, and in the absence of left-right mixing, they are
flavor-diagonal in the separate $LL$ and $RR$ sectors, even with the
presence of LFV in one or both.  Thus, in the absence of left-right
mixing, the Higgs coupling to sleptons is flavor-diagonal in the
6-dimensional space of left and right sleptons and does not contribute
to three-body decays.  Similarly, the $Z$ couples to the $LL$ and $RR$
sleptons separately, so in the absence of left-right mixing, its
couplings are also flavor-diagonal.  Note, however, there are no
photon-mediated diagrams at tree level, as the photon couples with
equal strength to the left- and right-handed sleptons, and so its
couplings are always flavor-diagonal.  Also note that in the case of
non-zero left-right mixing but no flavor mixing, the $H_a$ and $Z$
diagrams exist, but their contributions are only appreciable for
certain mass orderings.  Namely, only when neither of the two mass
eigenstates produced by left-right mixing posses kinematically allowed
two-body decays will the three-body $H_a$ and $Z$ contributions be
relevant.

The charge-flipping decay is unaffected by the Higgs and $Z$ boson
diagrams, but in the presence of flavor violation, the fermions in the
final state may be interchanged.  This implies a new interference
effect in the charge-flipping mode which is absent in the
flavor-conserving case.  In principle, there should also be a similar
interference term in the neutrino decay modes if neutrinos possess
Majorana masses, but such a term is negligible in the limit of zero
neutrino mass.

In summary, for charge-preserving decays to charged leptons of
different generations, the analysis of Ref.~\cite{Ambrosanio:1997bq}
may be straightforwardly modified to the general case.  The only
modification needed is the insertion of rotation matrices from flavor
eigenstates to gauge interaction eigenstates.  For charge-preserving
decays to charged leptons of the same generation, however, the Higgs
and $Z$ diagrams introduce new contributions to the decay width.
There are also new charge-preserving decays to neutrinos, mediated by
charginos and $Z$ bosons, and to quarks, mediated by Higgs and $Z$
bosons. Finally, in the case of the charge-flipping decays to
like-sign leptons, the daughter leptons may be interchanged and thus
the decay width must include an interference term between these two
diagrams; however, there are no Higgs and $Z$ contributions to this
mode and so there are no further interference terms.


\section{Analytic Results and Qualitative Observations}
\label{sec:widths}

Given an understanding of the qualitatively new features introduced by
flavor and left-right mixing described in \secref{diagrams}, we can
now calculate the resulting decay widths.  The full results are
lengthy and are given in the appendices.  In this section, we
summarize the results of our calculations and provide qualitative
insight into the relative importance of each contributing mode.

Our slepton interaction Lagrangian is
\begin{eqnarray}
\mathcal{L}_{\tilde{\ell}}^{\text{int}} &= & \left[ \tilde{\ell}_i^*
\overline{\tilde{\chi}}_a^0 \left( \beta_{aik}^{(1)} P_L +
\beta_{aik}^{(2)} P_R \right) \ell_k + \gamma_{aik} \tilde{\ell}_i^*
\overline{\tilde{\chi}}_a P_L \nu_k + \text{h.c.} \right] + i
\zeta^{(2)}_{ij} \left(\tilde{\ell}_i^* \partial_{\mu} \tilde{\ell}_j
- \tilde{\ell}_j \partial_{\mu} \tilde{\ell}_i^* \right) Z^{\mu}
\nonumber \\ 
&& + \sigma_{ij}^{(2)} \tilde{\ell}_i^* \tilde{\ell}_j h^0 +
\sigma_{ij}^{(3)} \tilde{\ell}_i^* \tilde{\ell}_j H^0 + i
\sigma_{ij}^{(4)} \tilde{\ell}_i^* \tilde{\ell}_j A^0 + i
\sigma_{ij}^{(5)} \tilde{\ell}_i^* \tilde{\ell}_j G^0 \ ,
\label{Lint}
\end{eqnarray}
where the coefficients $\beta$, $\gamma$, $\zeta$, and $\sigma$
contain gauge and Yukawa couplings along with the mixing matrix
elements for sleptons, Higgs bosons, neutralinos, and charginos.
These coefficients are defined in \appref{lagrangian}.  Here, the
indices are summed over the six slepton, three lepton and neutrino,
four neutralino, and two chargino mass eigenstates.  We have kept only
the terms relevant to the three-body decays in question.

Our lepton interaction Lagrangian is
\begin{eqnarray}
\mathcal{L}_{\ell}^{\text{int}} &=& \frac{y_k^{(\ell)}}{\sqrt{2}}
\overline{\ell}_k \left( h^0 \sin \theta_H - H^0 \cos \theta_H + i A^0
\gamma^5 \sin \beta - i G^0 \gamma^5 \cos \beta \right) \ell_k
\nonumber \\ 
&&+ \frac{ig}{2 \cos \theta_W} Z_{\mu} \left(1 - 2 \sin^2 \theta_W
\right) \overline{\ell}_k \gamma^{\mu} \ell_k \ .
\end{eqnarray}

The amplitudes for charge-preserving decays to charged leptons
$\tilde{\ell}_i^- \to \tilde{\ell}_j^- \ell_k^- \ell_m^+$ are
\begin{eqnarray}
\label{eqn:preservingMEstart}
\mathcal{M}_{\tilde{\chi}_a^0} &=& - i \overline{u} \left( p_k \right)
\left( \beta_{aik}^{1 *} P_R + \beta_{aik}^{2 *} P_L \right)
\frac{\left( \slam{p}_k - \slam{p}_i \right) + m_{\tilde{\chi}_a^0}}
{\left( p_k - p_i \right)^2 - m_{\tilde{\chi}_a^0}^2} \left(
\beta_{ajm}^1 P_L + \beta_{ajm}^2 P_R \right) v\left(p_m\right) \\
\mathcal{M}_Z &=& \frac{i g \zeta_{ji}^{(2)}}{ 2 \cos \theta_W}
\frac{1}{\left(p_i - p_j\right)^2 - m_Z^2} \overline{u}
\left(p_k\right) \left(\slashed{p}_i + \slashed{p}_j\right) \left(2
\sin^2 \theta_W - P_L\right) v \left(p_m\right) \delta_{km} \\
\mathcal{M}_{h^0} &=& \frac{- i \sigma_{ji}^{(2)} y_k^{(\ell)} \sin
\theta_H}{\sqrt{2}} \frac{1}{\left(p_i - p_j\right)^2 - m_{h^0}^2}
\overline{u} \left(p_k\right) v \left(p_m\right) \delta_{km} \\
\mathcal{M}_{H^0} &=& \frac{i \sigma_{ji}^{(3)} y_k^{(\ell)} \cos
\theta_H}{\sqrt{2}} \frac{1}{\left(p_i - p_j\right)^2 - m_{H^0}^2}
\overline{u} \left(p_k\right) v \left(p_m\right) \delta_{km} \\
\mathcal{M}_{A^0} &=& \frac{- i \sigma_{ji}^{(4)} y_k^{(\ell)} \sin
\beta}{\sqrt{2}} \frac{1}{\left(p_i - p_j\right)^2 - m_{A^0}^2}
\overline{u} \left(p_k\right) \gamma^5 v \left(p_m\right) \delta_{km}
\\
\mathcal{M}_{G^0} &=& \frac{i \sigma_{ji}^{(5)} y_k^{(\ell)} \cos
\beta}{\sqrt{2}} \frac{1}{\left(p_i - p_j\right)^2 - m_Z^2}
\overline{u} \left(p_k\right) \gamma^5 v \left(p_m\right) \delta_{km}
\ , \label{eqn:preservingMEend}
\end{eqnarray}
where the indices $i$, $j$, $k$, and $m$ correspond to the subscripts
specifying the mass eigenstates of initial and final state particles
in the decay mode $\tilde{\ell}_i^- \to \tilde{\ell}_j^- \ell_k^-
\ell_m^+$, and $\delta_{km}$ is the Kronecker delta function.  Here
and throughout the rest of this section we suppress these indices on
matrix elements and decay widths.

The overall decay width is given by
\begin{eqnarray}
\Gamma \left(\tilde{\ell}_i^- \to \tilde{\ell}_j^- \ell_k^-
\ell_m^+\right) & = &\frac{1}{64 \pi^3 m_i} \intps \sum_{\text{spins}}
|\mathcal{M}|^2 \\
& = &\frac{1}{64 \pi^3 m_i} \intps \sum_{\text{spins}} \left(
\sum_{a,b=1}^4 \mathcal{M}_{\tilde{\chi}_a^0}
\mathcal{M}_{\tilde{\chi}_b^0}^* + \mathcal{M}_Z \mathcal{M}_Z^* +
\sum_{a,b} \mathcal{M}_{H_a} \mathcal{M}_{H_b}^* \right. \nonumber \\
&& \left.  + 2 \text{ Re} \left[\sum_{b=1}^4 \mathcal{M}_Z
\mathcal{M}_{\tilde{\chi}_b^0}^* + \sum_{a} \sum_{b=1}^4
\mathcal{M}_{H_a} \mathcal{M}_{\tilde{\chi}_b^0}^* + \sum_b
\mathcal{M}_Z \mathcal{M}_{H_b}^* \right] \right) \\ 
& = &\Gamma_{\tilde{\chi}^0 \tilde{\chi}^0} + \delta_{km}
\left(\Gamma_{ZZ} + \Gamma_{HH} + 2 \Gamma_{Z \tilde{\chi}^0} + 2
\Gamma_{H \tilde{\chi}^0} + 2 \Gamma_{ZH} \right) \ ,
\end{eqnarray}
where $\intps$ is the integral over three-body phase space discussed
in \appref{phasespace} which includes a sum over fermion spins, and we
use an intuitively obvious notation for widths, so that, for example,
$\Gamma_{\tilde{\chi}^0 \tilde{\chi}^0}$ is the partial width from
$\sum_{a,b} \mathcal{M}_{\tilde{\chi}_a^0}
\mathcal{M}_{\tilde{\chi}_b^0}^*$.

To develop a qualitative understanding of which matrix elements are
typically dominant and sub-dominant, we first note that there is a
suppression to the Higgs and $Z$ modes from left-right slepton mixing.
In fact, the explicit Higgs and $Z$ couplings to sleptons depend on
the left-right slepton mixing squared,
\begin{equation}
\sigma_{ij}^{(a)}, \zeta_{ij}^{(2)} \sim 
\frac{m_{LR}^2}{m_{\tilde{\ell}}^2} \ .
\end{equation}
In particular, for $\mu$ not significantly larger than the slepton
mass, this is roughly the same order as Yukawa suppression.  We thus
expect that, at the matrix element level, the Higgs or $Z$ radiated
modes are generically suppressed by two powers of left-right mixing
relative to the neutralino and chargino modes.

Second, to simplify phase space factors, we assume a typical amount of
mass squared splitting $\Delta m_{\tilde{\ell}}^2 / m_{\tilde{\ell}}^2
\alt 5\% $; we also neglect lepton masses when they are sub-dominant.
Under these assumptions, the difference in propagator structure
between the neutralino mode and that of the Higgs and $Z$ modes
becomes apparent.  The neutralino propagator is inversely proportional
to $\left(p_k - p_i\right)^2 - m_{\tilde{\chi}_a^0}^2 \approx
m_{\tilde{\ell}_i}^2 - m_{\tilde{\chi}_a^0}^2$, while the Higgs ($Z$)
propagator is inversely proportional to $\left(p_i - p_j\right)^2 -
m_{H_a,Z}^2 \approx m_{\tilde{\ell}_i}^2 - m_{\tilde{\ell}_j}^2 -
m_{H_a,Z}^2$, which simply reduces to $- m_{H_a,Z}^2$ if the slepton
masses are not too far above $m_Z$.  Thus, in models where the
lightest neutralino and light slepton masses are close, the lightest
neutralino pole contribution will be enhanced over the Higgs and $Z$
mass-suppressed contributions.  Conversely, as the slepton and
neutralino mass scale grows, the Higgs and $Z$ contributions will drop
off more slowly than the neutralino contribution.

The three-body decays with two neutrinos or two quarks have much the
same form as above with minor changes (refer to the appendices for
details).  These changes are as follows: For the neutrino mode,
$\beta_{aik}^1 \to \gamma_{aik}$ and $\beta_{aik}^2 \to 0$, and the
lepton masses are set to zero, hence simplifying the phase space
calculation and removing the off-shell Higgs contribution.  For the
quark modes, there is no off-shell fermion intermediary, and the Higgs
and $Z$ modes are only modified with adjusted couplings and quark
masses.  The quark modes are, however, enhanced by the color factor
and a sum over light flavors.  Because of these changes, however, all
of these modes are suppressed relative to the di-lepton mode by
left-right mixing.  The quark modes only have contributions from the
Higgs and $Z$ diagrams, and thus are suppressed by left-right mixing
as noted above.  In the majority of models with a light slepton NLSP,
the $\tilde{\ell}_R$ gauge eigenstates are generically lighter than
the $\tilde{\ell}_L$ gauge eigenstates; for such a model the neutrino
modes are suppressed since the neutrino only couples to
$\tilde{\ell}_L$ gauge eigenstates while the light sleptons are
primarily $\tilde{\ell}_R$ states.  Indeed, in such a model
\begin{equation}
\gamma_{aik} \sim \left( \frac{m_{LR}^2}{m_{\tilde{\ell}}^2}
\right)^{\frac{1}{2}} \ ,
\end{equation}
so the neutrino and quark decay widths are all suppressed by $\left(
m_{LR}^2 / m_{\tilde{\ell}}^2 \right)^2$ relative to the dilepton
modes.

Finally, for the charge-flipping di-lepton decay, the matrix
element is given by
\begin{eqnarray}
\label{eqn:flippingMEstart}
\mathcal{M}_{\tilde{\chi}_a^0}^{(1)} &=& -i \overline{u}
\left(p_k\right) \left( \beta_{aik}^{1 *} P_R + \beta_{aik}^{2 *} P_L
\right) \frac{\left( \slashed{p}_k - \slashed{p}_i \right) +
m_{\tilde{\chi}_a^0}}{\left(p_i - p_k\right)^2 -
m_{\tilde{\chi}_a^0}^2} \left( \beta_{ajm}^{2*} P_L + \beta_{ajm}^{1*}
P_R \right) v\left(p_m\right) \\ 
\mathcal{M}_{\tilde{\chi}_a^0}^{(2)} &=& - i \overline{u}
\left(p_k\right) \left( \beta_{aim}^{1 *} P_R + \beta_{aim}^{2 *} P_L
\right) \frac{\left(\slashed{p}_k - \slashed{p}_i \right) +
m_{\tilde{\chi}_a^0}}{\left(p_i - p_k\right)^2 -
m_{\tilde{\chi}_a^0}^2} \left( \beta_{ajk}^{2*} P_L + \beta_{ajk}^{1*}
P_R \right) v\left(p_m\right) \\ 
\mathcal{M}_{\tilde{\chi}_a^0} &=&
\mathcal{M}_{\tilde{\chi}_a^0}^{(1)} -
\mathcal{M}_{\tilde{\chi}_a^0}^{(2)} \ ,
\label{eqn:flippingMEend}
\end{eqnarray}
where the negative sign comes from Fermi statistics.  The decay width
is then
\begin{equation}
\Gamma \left(\tilde{\ell}_i^- \to \tilde{\ell}_j^+ \ell_k^-
\ell_m^-\right) = C_{km} \left( \Gamma_{11} + \Gamma_{22} - 2
\Gamma_{21} \right) \ ,
\end{equation}
with 
\begin{equation}
\Gamma_{ij} = \frac{1}{64 \pi^3 m_i} \intps \sum_{\text{spins}}
\sum_{a,b=1}^4 \text{ Re} \left[ \mathcal{M}_a^{(i)}
\mathcal{M}_b^{(j)*} \right],
\end{equation}
and $C_{km}$ is a phase space factor which has a value of 1 for two
outgoing leptons of different flavor and 1/2 for two outgoing leptons
of the same flavor due to indistinguishable particle statistics.
$\Gamma_{11}$ and $\Gamma_{22}$ both have the same basic form as the
result from \cite{Ambrosanio:1997bq}, where the only change is the
insertion of flavor-mixing coefficients and, for $\Gamma_{22}$,
interchange of $\ell_k$ and $\ell_m$ between the two terms; the new
$\Gamma_{21}$ is presented in \appref{gamma21}.  We find that this
charge-flipping decay width is of the same order of magnitude as the
charge-preserving di-lepton mode, though the flavor structure is
markedly different: this is discussed in the next section.


\section{Two-Slepton Mixing: Illustrative Examples}
\label{sec:plots}

To validate our results and investigate their phenomenological
implications, we examine some simple cases of two-slepton mixing.  In
these examples we consider spectra with fairly degenerate sleptons and
lightest neutralino.  The slepton-neutralino degeneracy is motivated
by simple gauge-mediated supersymmetry breaking scenarios with not too
many messengers.  Larger splittings are, of course, possible if there
are many messengers or in other frameworks, such as minimal
supergravity.  The sleptons are taken fairly degenerate so that large
mixing angles are consistent with low energy constraints and so may be
considered.  We note, however, that these examples are merely
illustrative, and our results are valid in any chosen framework with
arbitrary mass splittings.

\begin{figure}[tb]
\includegraphics[width=8cm]{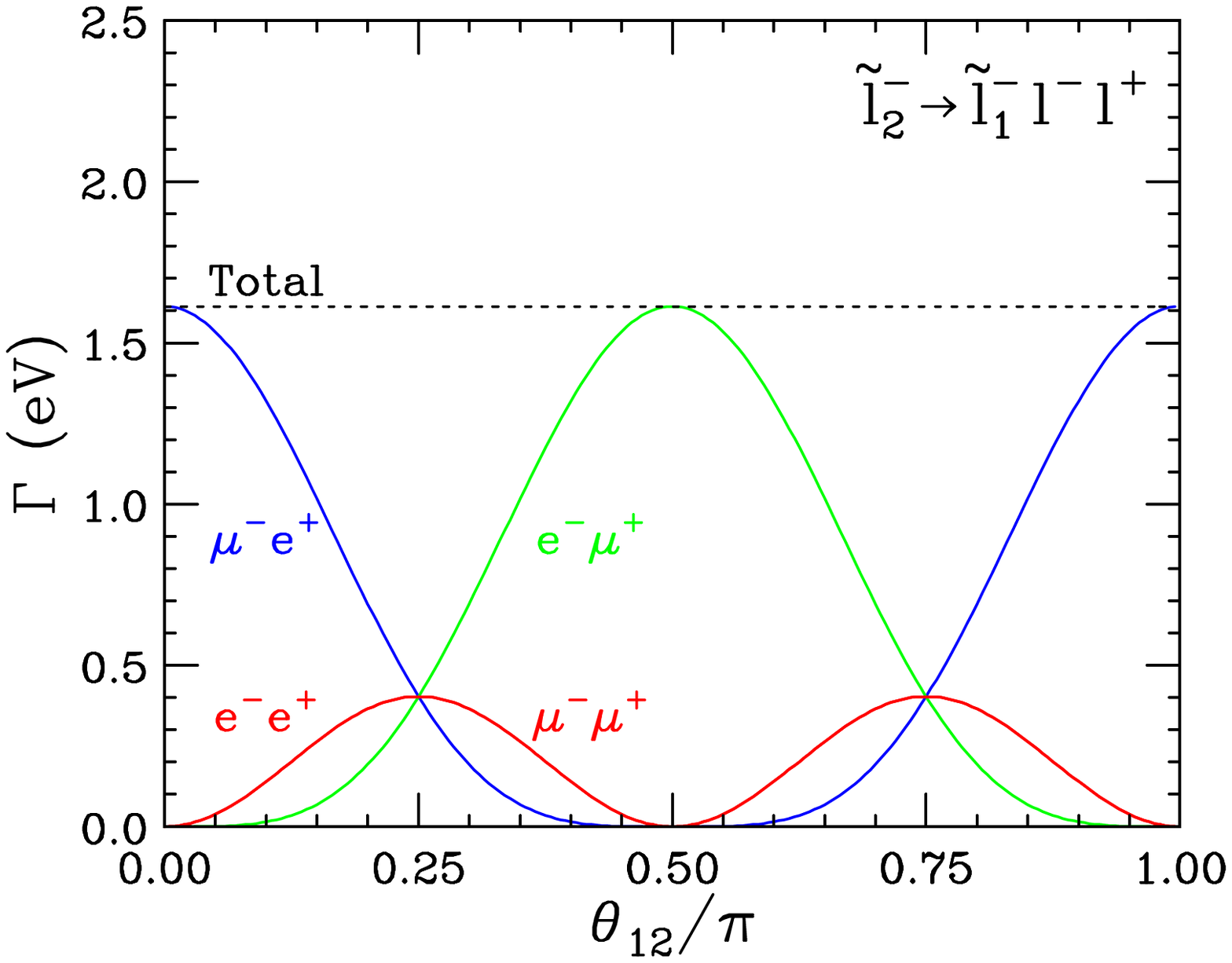}
\hspace*{.01in}
\includegraphics[width=8cm]{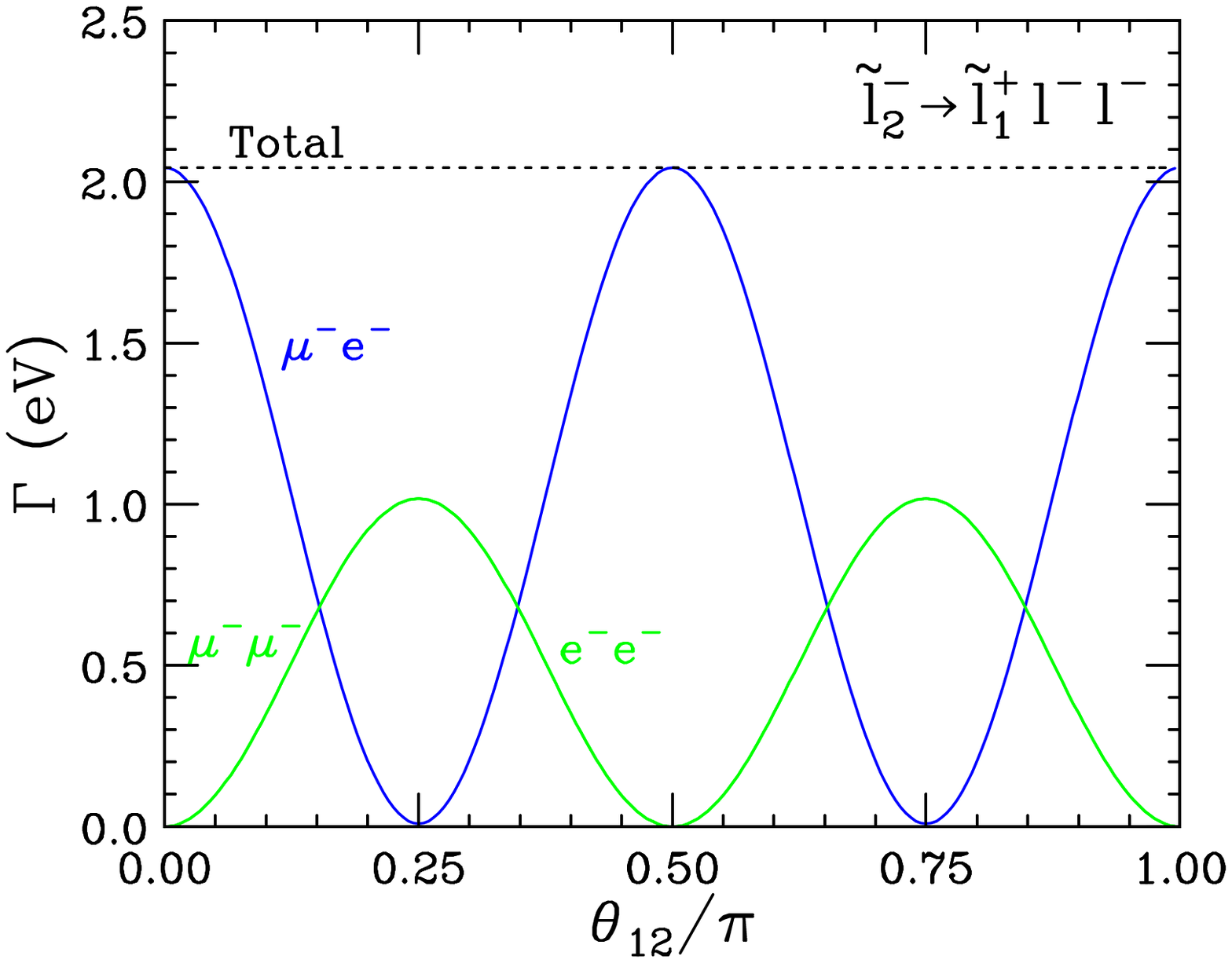}
\caption{Di-lepton decay widths as a function of mixing angle between
smuon and selectron with left-right mixing neglected.  The left plot
shows the charge-preserving channel, the right plot shows the
charge-flipping channel.  In both plots the leptons are taken to be
massless, $m_{\tilde{\ell}_1} = 100$ GeV, $m_{\tilde{\ell}_2} = 105$
GeV, and $m_{\tilde{\chi}_1^0} = 110$ GeV.}
\label{fig:smuse}
\end{figure}

First, we consider $\tilde{e}_R-\tilde{\mu}_R$ mixing parameterized by
\begin{equation}
\left( \begin{array}{c} 
\tilde{\ell}_1 \\ \tilde{\ell}_2 
\end{array} \right)
= \left( \begin{array}{cc} \cos \theta_{12} & \sin \theta_{12} \\
- \sin \theta_{12} & \cos \theta_{12}
\end{array} \right)
\left( \begin{array}{c} 
\tilde{e} \\ \tilde{\mu}
\end{array} \right) .
\end{equation}  
\figref{smuse} shows the flavor-violating decay widths
$\tilde{\ell}_2^- \to \tilde{\ell}_1^\pm \ell^- \ell^\mp$ with
electron and/or muon leptons as a function of the mixing angle
$\theta_{12}$.  Here both leptons are taken to be explicitly massless
and left-right slepton mixing is set to zero, thereby cutting off the
Higgs and $Z$ modes.  $\tilde{\ell}_1$ and $\tilde{\ell}_2$ are
assigned masses of 100 GeV and 105 GeV, respectively, and the lightest
neutralino is given a mass of 110 GeV (heavier neutralino
contributions are small in this case).

The left plot in \figref{smuse}, which shows the decay widths for the
charge-preserving channel, demonstrates the typical structure of
two-slepton mixing: all decay widths are at most $\pi$ periodic, the
sleptons interchange roles ($\mu^- \to e^-$ and $e^+ \to \mu^+$) at
$\theta_{12} = \pi/2$, the $e^- e^+$ and $\mu^- \mu^+$ modes are equal
at all mixing angles, and at $\theta_{12} = \pi/4$ all decay widths
are equal.  In contrast, the right plot in \figref{smuse}, which shows
the charge-flipping channel, demonstrates a different flavor
structure.  Since the $e^- \mu^-$ and $\mu^- e^-$ modes in general
contribute to the same decay width, all widths are $\pi/2$ periodic.
Also, the $e^- e^-$ and $\mu^- \mu^-$ modes are again equal at all
mixing angles.  In addition, note that the $\mu^- e^-$ mode drops to
zero at $\theta_{12} = \pi/4$, because the charge-flipping mode has
two diagrams at tree level which cancel at $\theta_{12} = \pi/4$ for
the $\mu^- e^-$ mode.  Then, as expected, the $e^- e^-$ and $\mu^-
\mu^-$ modes have decay widths that are half the total width at
$\theta_{12} = \pi/4$.

\begin{figure}[tb]
\includegraphics[width=8cm]{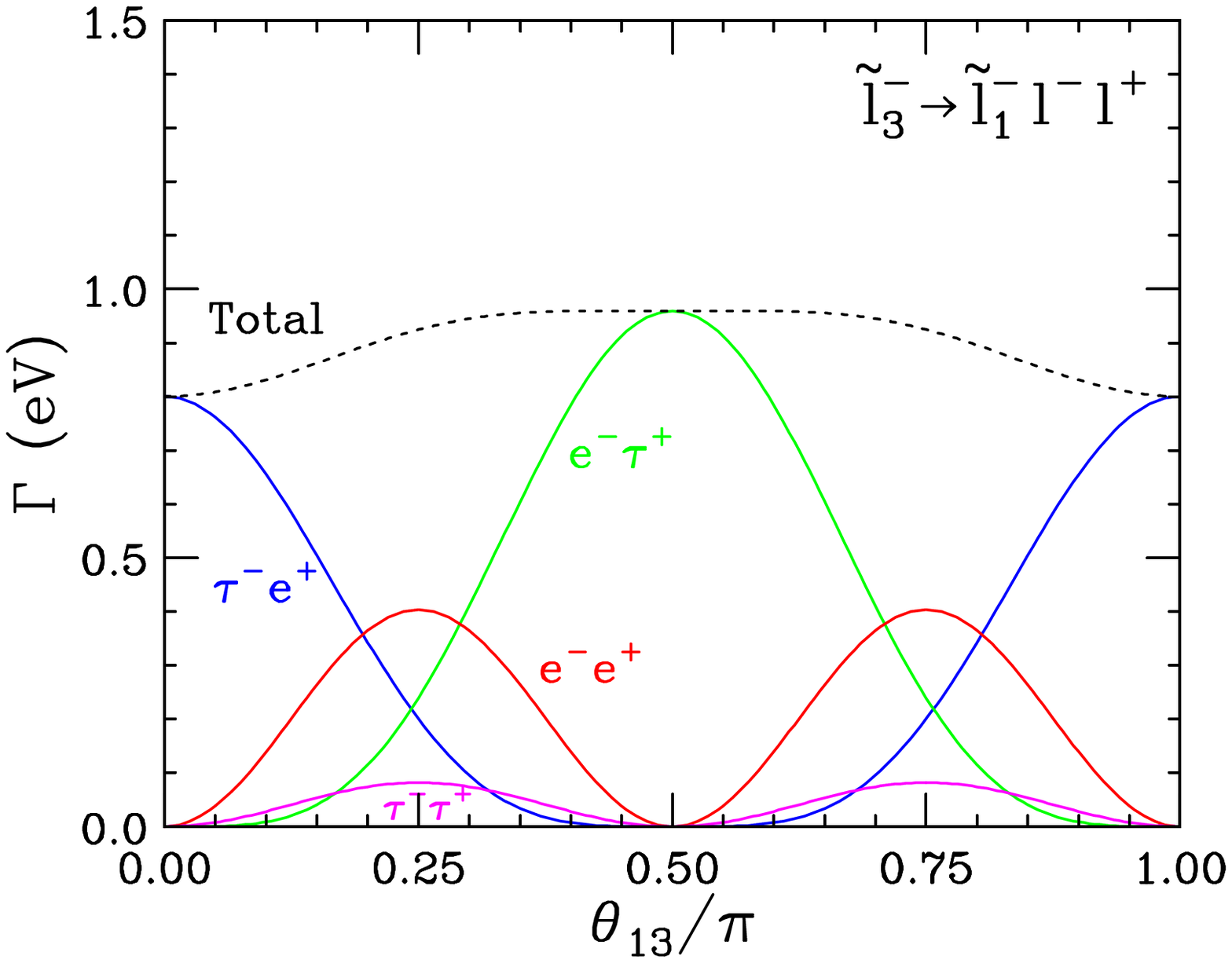}
\hspace*{.01in}
\includegraphics[width=8cm]{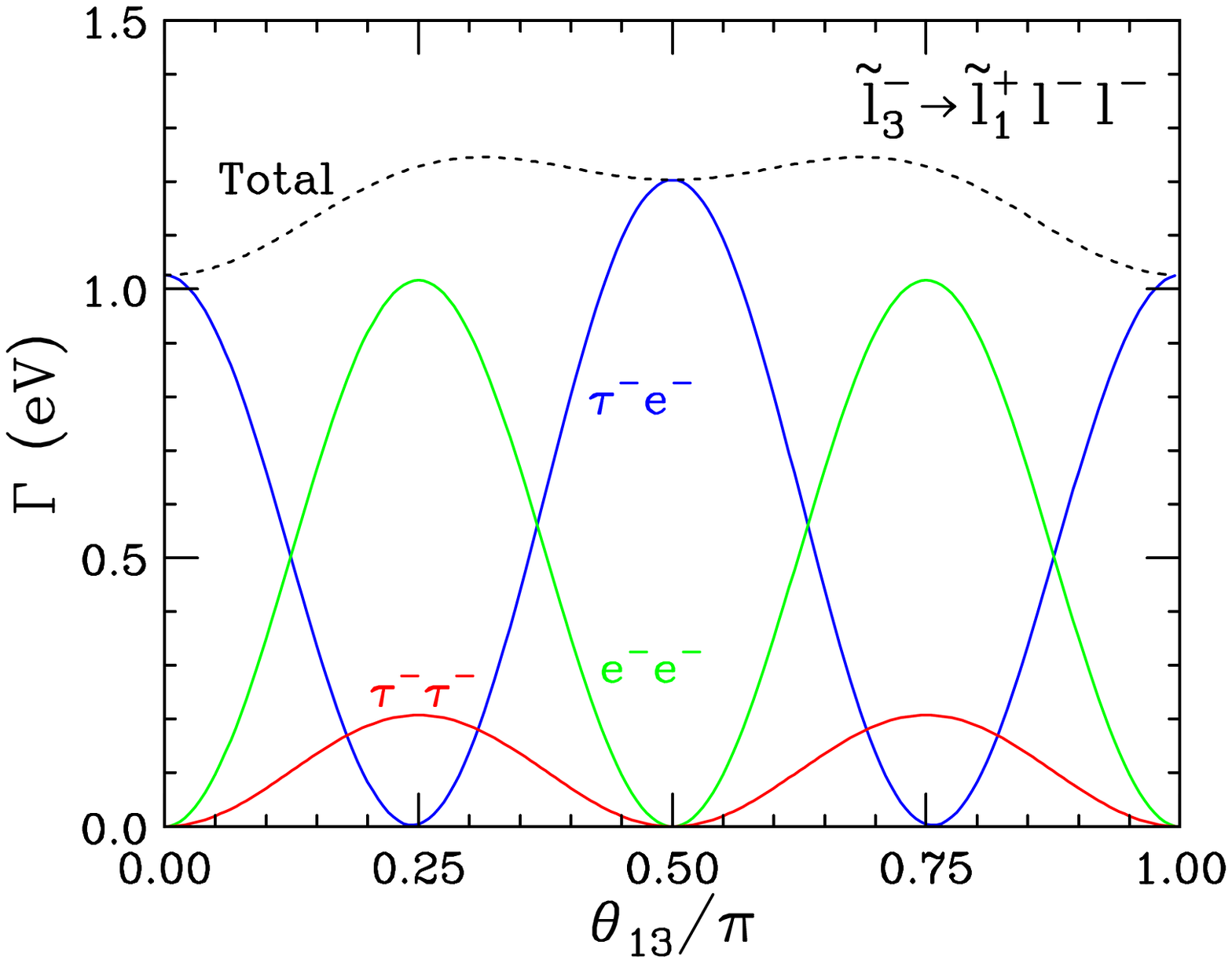}
\caption{Di-lepton decay widths as a function of mixing angle between
stau and selectron with left-right mixing neglected..  The left plot
shows the charge-preserving channel, the right plot shows the
charge-flipping channel.  Here the leptons are both massive,
$m_{\tilde{\ell}_1} = 100$ GeV, $m_{\tilde{\ell}_3} = 105$ GeV, and
$m_{\tilde{\chi}_1^0} = 110$ GeV.}
\label{fig:stause}
\end{figure}

Next we consider $\tilde{e}_R-\tilde{\tau}_R$ mixing, with massive
leptons and a mixing angle $\theta_{13}$ given by
\begin{equation}
\left( \begin{array}{c} 
\tilde{\ell}_1 \\ \tilde{\ell}_3 
\end{array} \right)
= \left( \begin{array}{cc} \cos \theta_{13} & \sin \theta_{13} \\
- \sin \theta_{13} & \cos \theta_{13}
\end{array} \right)
\left( \begin{array}{c} 
\tilde{e} \\ \tilde{\tau} 
\end{array} \right) .
\end{equation}
\Figref{stause} shows the widths for $\tilde{\ell}_3^- \to
\tilde{\ell}_1^\pm \ell^- \ell^\mp$, where the leptons are electrons
and/or taus.  $\tilde{\ell}_1$ and $\tilde{\ell}_3$ have masses of 100
GeV and 105 GeV, and lightest neutralino mass is again 110 GeV.
Left-right mixing is again neglected.

The first notable feature in \Figref{stause} is the separation between
the $e^- e^\pm$ and $\tau^- \tau^\pm$ modes: the $e^- e^\pm$ mode is
almost unchanged numerically from the previous case, while the $\tau^-
\tau^\pm$ mode is suppressed by phase space constriction, as expected.
Likewise, the $\tau^- e^\pm$ and $e^- \tau^\pm$ modes are also
suppressed relative to the massless case, but certainly less
suppressed than the $\tau^- \tau^\pm$ mode.  Note that the total decay
width is no longer constant because of the nonzero tau mass.

\section{Conclusions and Implementation in SPICE}
\label{sec:conclusion}

We have determined the decay widths of three-body slepton decays,
including both ``charge-preserving'' processes $\tilde{\ell}^- \to
\tilde{\ell}^- \ell^- \ell^+$ and ``charge-flipping'' processes
$\tilde{\ell}^- \to \tilde{\ell}^+ \ell^- \ell^-$, in the presence of
arbitrary lepton flavor violation (LFV) and arbitrary left-right
mixing.  Such processes are particularly relevant in scenarios with a
gravitino LSP and a slepton NLSP, where they are typically the
dominant decay of some of the non-NLSP sleptons and are present as the
last step in many SUSY cascade decays.

Arbitrary $6\times 6$ slepton mixing leads to many new diagrams and
new decade modes, as well as new interference effects.  Our results
are fully general, but we have illustrated them for two simple cases
with 2-generation LFV.  The branching ratios to new modes may be
significant, even accounting for low-energy constraints on LFV.  The
new modes may substitute final state electrons and muons for taus,
potentially providing more obvious signals and better prospects for
precision mass measurements.  In addition, the measurement of these
branching ratios may help pin down the supersymmetric flavor
parameters and shed light on the SM flavor problem.  We note that our
illustrative examples have assumed fairly small slepton mass
splittings so that large mixing angles are consistent with low energy
constraints.  It is also possible, of course, that the splittings are
larger and the mixing angles are smaller.  The calculations presented
here are valid also in those cases, and it would also be very
interesting to determine whether such small mixings may be established
through collider studies.

The full calculation presented here is lengthy, but has been
incorporated into {\tt SPICE}: Simulation Package for Including Flavor
in Collider Events, a publicly available code. SPICE takes as input a
flavor-conserving model, such as minimal gauge-mediated supersymmetry
breaking, and arbitrary lepton flavor-violating parameters, and
generates the full supersymmetric spectrum, complete with
flavor-violating branching ratios.  The three-body decays discussed
here are included in {\tt SPICE}.  Details on obtaining and running
{\tt SPICE} are described in the {\tt SPICE} user
manual~\cite{Engelhard:SPICE}.


\section*{Acknowledgements}
\label{sec:acknowledgements}

We are grateful to Y.~Nir for many helpful discussions and a careful
reading of the manuscript.  The work of JLF, DS, and FY was supported
in part by NSF Grant No.~PHY--0239817 and No.~PHY--0653656 and the Alfred
P.~Sloan Foundation.  IG thanks the UC Irvine particle theory group
for their hospitality while this work was in progress.  This research was
supported in part by the United States-Israel Binational Science
Foundation (BSF) under Grant No.~2006071.  The research of IG and YS
is also supported by the Israel Science Foundation (ISF) under Grant
No.~1155/07.


\appendix

\section{Lagrangian}
\label{app:lagrangian}

As given in \eqref{Lint}, the relevant flavor-general interactions for
this study are given by the Lagrangian
\begin{eqnarray}
\mathcal{L}_{\tilde{\ell}}^{\text{int}} &= & \left[ \tilde{\ell}_i^*
\overline{\tilde{\chi}}_a^0 \left( \beta_{aik}^{(1)} P_L +
\beta_{aik}^{(2)} P_R \right) \ell_k + \gamma_{aik} \tilde{\ell}_i^*
\overline{\tilde{\chi}}_a P_L \nu_k + \text{h.c.} \right] + i
\zeta^{(2)}_{ij} \left(\tilde{\ell}_i^* \partial_{\mu} \tilde{\ell}_j
- \tilde{\ell}_j \partial_{\mu} \tilde{\ell}_i^* \right) Z^{\mu}
\nonumber \\ 
&& + \sigma_{ij}^{(2)} \tilde{\ell}_i^* \tilde{\ell}_j h^0 +
\sigma_{ij}^{(3)} \tilde{\ell}_i^* \tilde{\ell}_j H^0 + i
\sigma_{ij}^{(4)} \tilde{\ell}_i^* \tilde{\ell}_j A^0 + i
\sigma_{ij}^{(5)} \tilde{\ell}_i^* \tilde{\ell}_j G^0 \ .
\end{eqnarray}
The coefficients are
\begin{eqnarray}
\beta_{aik}^{(1)} &=& \frac{1}{\sqrt{2}}
\left( g O_{2, a}^* + g' O_{1, a}^* \right) U_{k, i}^{(\tilde{\ell})*} 
- y_{k}^{(\ell)} O_{3, a}^* U_{k + 3, i}^{(\tilde{\ell})*} \\
\beta_{aik}^{(2)} &=& - \sqrt{2} O_{1, a} g' U_{k + 3, i}^{(\tilde{\ell})*}
- y_{k}^{(\ell)} O_{3, a} U_{k, i}^{(\tilde{\ell})*} \\ 
\gamma_{aik} &=& \left\{ \begin{array}{ll}
-g \cos \theta_L \, U_{k, i}^{(\tilde{\ell})*} +
y_{k}^{(\ell)} \sin \theta_L \, U_{k + 3, i}^{(\tilde{\ell})*}
& \qquad a = 1 \\
g \sin \theta_L \, U_{k, i}^{(\tilde{\ell})*} +
y_{k}^{(\ell)} \cos \theta_L \, U_{k + 3, i}^{(\tilde{\ell})*}\,
& \qquad a = 2 \\
\end{array} \right. \\
\zeta^{(2)}_{ij} & = & \frac{g}{2 \cos \theta_W} 
\left[ U_{c, i}^{(\tilde{\ell})*} U_{c, j}^{(\tilde{\ell})}
- 2 \sin^2 \theta_W \delta_{ij} \right] \\
\nonumber \sigma_{ij}^{(2)} &=& - \left[ \left( \frac{g m_W}{2} (1 -
\tan^2 \theta_W) \sin (\theta_H + \beta) - \frac{g m_{\ell_c}^2 \sin
\theta_H}{ m_W \cos \beta} \right) U_{c,i}^{(\tilde{\ell})*}
U_{c,j}^{(\tilde{\ell})} \right. \\
& & \quad + \left( g m_W \tan^2 \theta_W \sin (\theta_H + \beta) -
\frac{g m_{\ell_c}^2 \sin \theta_H}{ m_W \cos \beta} \right)
U_{c+3,i}^{(\tilde{\ell})*} U_{c+3,j}^{(\tilde{\ell})} \nonumber \\
& & \quad \left.  - \frac{g m_{\ell_c}}{2 m_W \cos \beta} \left(\mu
\cos \theta_H + A_c^{\tilde{\ell}} \sin \theta_H \right)
\left (U_{c,i}^{(\tilde{\ell})*} U_{c+3,j}^{(\tilde{\ell})} +
U_{c+3,i}^{(\tilde{\ell})*} U_{c,j}^{(\tilde{\ell})}\right) \right] \\
\nonumber \sigma_{ij}^{(3)} &=& \left[ \left( \frac{g m_W}{2} (1 -
\tan^2 \theta_W) \cos (\theta_H + \beta) - \frac{g m_{\ell_c}^2 \cos
\theta_H}{ m_W \cos \beta} \right) U_{c,i}^{(\tilde{\ell})*}
U_{c,j}^{(\tilde{\ell})} \right. \\
& & \quad + \left( g m_W \tan^2 \theta_W \cos (\theta_H + \beta) -
\frac{g m_{\ell_c}^2 \cos \theta_H}{ m_W \cos \beta} \right)
U_{c+3,i}^{(\tilde{\ell})*} U_{c+3,j}^{(\tilde{\ell})} \nonumber \\
& &\quad \left.  + \frac{g m_{\ell_c}}{2 m_W \cos \beta} \left(\mu
\sin \theta_H - A_c^{\tilde{\ell}} \cos \theta_H \right)
\left(U_{c,i}^{(\tilde{\ell})*} U_{c+3,j}^{(\tilde{\ell})} +
U_{c+3,i}^{(\tilde{\ell})*} U_{c,j}^{(\tilde{\ell})}\right) \right] \\
\sigma_{ij}^{(4)} &=& \frac{g m_{\ell_c}}{ 2 m_W} (\mu +
A_c^{\tilde{\ell}} \tan \beta) \left( U_{c, i}^{(\tilde{\ell})*} U_{c
+ 3, j}^{(\tilde{\ell})} - U_{c + 3, i}^{(\tilde{\ell})*} U_{c,
j}^{(\tilde{\ell})} \right) \\
\sigma_{ij}^{(5)} &=& \frac{g m_{\ell_c}}{ 2 m_W} (\mu \tan \beta +
A_c^{\tilde{\ell}}) \left( U_{c, i}^{(\tilde{\ell})*} U_{c + 3,
j}^{(\tilde{\ell})} - U_{c + 3, i}^{(\tilde{\ell})*} U_{c,
j}^{(\tilde{\ell})} \right).
\end{eqnarray}
$U_{c, a}^{(\tilde{\ell})}$ and $U_{c, a}^{(\tilde{\nu})}$ are the
slepton and sneutrino mixing matrices, respectively. These relate the
slepton and sneutrino gauge eigenstates,
\begin{eqnarray}
\tilde{\ell}_\alpha & = & \left( \tilde{e}_L, \tilde{\mu}_L,
\tilde{\tau}_L, \tilde{e}_R, \tilde{\mu}_R, \tilde{\tau}_R \right)^T
\\
\tilde{\nu} & = & \left(\tilde{\nu}_e, \tilde{\nu}_\mu,
\tilde{\nu}_\tau \right)^T
\end{eqnarray}
to the mass eigenstates $\tilde{\ell}_i, \tilde{\nu}_i$ through the
relations $\tilde{\ell}_i = U_{i \alpha}^{(\tilde{\ell})\dagger}
\tilde{\ell}_\alpha$ and $\tilde{\nu}_i = U_{i
\alpha}^{(\tilde{\nu})\dagger} \tilde{\nu}_\alpha$.  The mass
eigenstates are defined in order of increasing mass.  The neutralino
gauge eigenstates, $\tilde{\psi}^0 = \left( - i \tilde{B}, - i
\tilde{W}, \tilde{\psi}_d^0, \tilde{\psi}_u^0 \right)^T,$ are related
to the mass eigenstates $\chi_i^0$ by $\chi_i^0 = O_{i
\alpha}^{\dagger} \tilde{\psi}_\alpha^0$.  Similarly, the mixing of
the negative charginos is
\begin{equation}
\left( \begin{array}{c} \chi_1^- \\ \chi_2^- \end{array} \right) =
\left( \begin{array}{cc}
\cos \theta_L & \sin \theta_L \\ - \sin \theta_L & \cos \theta_L
\end{array} \right)
\left( \begin{array}{c} - i \tilde{W}^- \\ \tilde{\psi}_d^- \end{array} 
\right).
\end{equation}
Finally, the neutral Higgs doublet is given by 
\begin{equation}
\left( \begin{array}{c} 
H_u^0 \\ H_d^0 \\ 
\end{array} \right)
= \frac{1}{\sqrt{2}} 
\left( \begin{array}{c} 
v_u \\ v_d \\
\end{array} \right)
+ \frac{1}{\sqrt{2}} 
R_{\theta_H} 
\left( \begin{array}{c} 
h^0 \\ H^0 \\
\end{array} \right)
+ \frac{i}{\sqrt{2}} 
R_\beta 
\left( \begin{array}{c} 
G^0 \\ A^0 \\
\end{array} \right)\, ,
\end{equation}
where 
\begin{equation}
R_{\theta_H} =
\left( \begin{array}{cc}
  \cos \theta_H & \sin \theta_H \\
- \sin \theta_H & \cos \theta_H \\
\end{array} \right)
\quad \text{and} \quad
R_\beta =
\left( \begin{array}{cc}
  \sin \beta & \cos \beta \\
- \cos \beta & \sin \beta \\
\end{array} \right) \ .
\end{equation}

These coefficients are consistent with those of
Ref.~\cite{Engelhard:SPICE}, where a more complete discussion may be
found.


\section{Phase Space Integration}
\label{app:phasespace}

In the three-body decays $\tilde{\ell}_i \to \tilde{\ell}_j f_k f_m$,
we label the initial and final state particle masses $m_i$, $m_j$,
$m_k$, and $m_m$.  We will maintain these indices throughout the
appendix, with $i$ corresponding to the parent slepton, $j$ to the
daughter slepton, $k$ to the daughter fermion with the same sign as the
parent slepton, and $m$ to the other daughter fermion.  The decay widths
are the squared matrix elements integrated over 3-body phase space:
\begin{eqnarray}
\nonumber \Gamma = \frac{1}{64 \pi^3 m_i} \intps \left| {\cal M}
\right|^2 & \equiv & \frac{1}{64 \pi^3 m_i} \int_{E_k^-}^{E_k^+} dE_k
\int_{E_m^-}^{E_m^+} dE_m \left| {\cal M} \right|^2 \\
& = & \frac{m_i}{256 \pi^3} \int_{x_-}^{x_+} dx \int_{y_-}^{y_+} dy
\left| {\cal M} \right|^2 \ ,
\end{eqnarray}
where the dimensionless quantities 
\begin{equation}
x = \frac{2 E_k}{m_i} \quad \text{and} \quad y = \frac{2
E_m}{m_i}
\end{equation}
have limits
\begin{eqnarray}
x_+ &=& 1 + r_k^2 - \left( r_j + r_m \right)^2 \\
x_- &=& 2 r_k \\
y_\pm &=& \frac{1}{2 \left(1 - x + r_k^2\right)} 
\left[\left(2 - x\right)\left(R^2 - x\right) \pm \sqrt{x^2 - 4 r_k^2} \,  
\lambda^{\frac{1}{2}} \!
\left(1 - x + r_k^2, r_m^2, r_j^2 \right) \right]\, ,
\end{eqnarray}
and we have defined
\begin{eqnarray}
r_j &=& \frac{m_j}{m_i} \quad 
r_k = \frac{m_k}{m_i} \quad 
r_m = \frac{m_m}{m_i} \\
R^2 &=& 1 - r_j^2 + r_k^2 + r_m^2 \\
\lambda \left(x,y,z\right) &=& x^2 + y^2 + z^2 - 2xy - 2 xz - 2yz \ .
\end{eqnarray}
Throughout this work, we implicitly assume that the matrix element squared
has been averaged and summed over initial and final state spins,
respectively.

The squared matrix element has the form
\begin{equation}
\left| {\cal M} \right| ^2
= \sum_{a, b} {\cal M}_a {\cal M}_b^* \ ,
\end{equation}
where ${\cal M}_a$ and ${\cal M}_b$ are matrix elements corresponding
to Feynman diagrams with intermediate particle $a$ and $b$.  To
evaluate the decay widths, we choose to integrate over $y$
analytically.  The decay widths are then written in terms of
dimensionless integrals over $x$, which are performed numerically.  

In integrating over $y$, it is convenient to note that all terms
${\cal M}_a {\cal M}_b^*$ have a numerator that is either constant or
linear in $y$ and a denominator that is proportional to the product of
two propagators, each of which is either of the form $\left(1 - x +
r_k^2 - r_{a,b}^2\right)^{-1}$ or $\left(x + y - 1 + r_j^2 -
r_{a,b}^2\right)^{-1}$, where $r_{a,b} = m_{a,b}/m_i$.  We therefore
define the following dimensionless functions of $x$:
\begin{eqnarray}
f \left(x,r_a,r_b\right) &=& \int_{y_-}^{y_+} \frac{ y \, dy}{ \left(1
- x + r_k^2\right)\left(1 - x + r_k^2 - r_a^2\right)\left(1 - x +
r_k^2 - r_b^2\right)} \nonumber \\ 
&=& \frac{ k\left(x,r_b\right)} { \left(1 - x + r_k^2 -
r_a^2\right)\left(1 - x + r_k^2\right)} \\ 
f^{(1)} \left(x,r_a,r_b\right) &=& \int_{y_-}^{y_+} \frac{ dy}{
\left(1 - y + r_m^2 - r_a^2\right)\left(1 - x + r_k^2 - r_b^2\right)}
\nonumber \\ 
&=& - \frac{ l\left(x, r_a^2 - 1 - r_m^2\right)} { \left(1 - x + r_k^2
- r_b^2\right)} \\ 
f^{(2)} \left(x,r_a,r_b\right) &=& \int_{y_-}^{y_+} \frac{ y \, dy} {
\left(1 - y + r_m^2 - r_a^2\right)\left(1 - x + r_k^2 - r_b^2\right)}
\nonumber \\ 
&=& \left(1 + r_m^2 - r_a^2\right) f^{(1)}\left(x,r_a,r_b\right) -
k\left(x, r_b\right) \\ 
g^{(1)} \left(x,r_a,r_b\right) &=& \int_{y_-}^{y_+} \frac{ dy} {
\left(x + y - 1 + r_j^2 - r_a^2\right)\left(1 - x + r_k^2 -
r_b^2\right)} \nonumber \\ 
&=& \frac{ l\left(x, x - 1 +r_j^2 - r_a^2\right)} { \left(1 - x +
r_k^2 - r_b^2\right)} \\ 
g^{(2)} \left(x,r_a,r_b\right) &=& \int_{y_-}^{y_+} \frac{ y \, dy} {
\left(x + y - 1 + r_j^2 - r_a^2\right)\left(1 - x + r_k^2 -
r_b^2\right)} \nonumber \\ 
&=& k\left(x, r_b\right) + \left(1 - x - r_j^2 + r_a^2\right)
g^{(1)}\left(x,r_a,r_b\right) \\ 
h^{(1)} \left(x,r_a,r_b\right) &=& \int_{y_-}^{y_+} \frac{ dy} {
\left(x + y - 1 + r_j^2 - r_a^2\right)\left(x + y - 1 + r_j^2 -
r_b^2\right)} \nonumber \\ 
&=& \frac{ l\left(x, x - 1 +r_j^2 - r_a^2\right) - l\left(x, x - 1
+r_j^2 - r_b^2\right)}{ \left(r_a^2 - r_b^2\right)} \\ 
h^{(2)} \left(x,r_a,r_b\right) &=& \int_{y_-}^{y_+} \frac{ y \, dy} {
\left(x + y - 1 + r_j^2 - r_a^2\right)\left(x + y - 1 + r_j^2 -
r_b^2\right)} \nonumber \\ 
&=& l\left(x, x - 1 +r_j^2 - r_b^2\right) + \left(1 - x - r_j^2 +
r_a^2\right) h^{(1)}\left(x,r_a,r_b\right) \\ 
i^{(1)} \left(x,r_a\right) &=& \int_{y_-}^{y_+} \frac{ dy} { \left(x +
y - 1 + r_j^2 - r_a^2\right)^2} \nonumber \\ 
&=& \frac{ y_+ - y_-}{ \left(y_+ + x - 1 + r_j^2 - r_a^2\right)
\left(y_- + x - 1 + r_j^2 - r_a^2\right)}\\ 
i^{(2)} \left(x,r_a\right) &=& \int_{y_-}^{y_+} \frac{ y \, dy} {
\left(x + y - 1 + r_j^2 - r_a^2\right)^2} \nonumber \\ 
&=& l\left(x, x - 1 +r_j^2 - r_a^2\right) + \left(1 - x - r_j^2 +
r_a^2\right) i^{(1)}\left(x,r_a\right)\ ,
\end{eqnarray}
where
\begin{eqnarray}
k\left(x, r_b\right) & = & \frac{ y_+ - y_-}
{ 1 - x + r_k^2 - r_b^2} \\
l\left(x, z\right) & = & \ln \left| \frac{ y_+ + z}
{ y_- + z} \right| \ .
\end{eqnarray}

These functions will appear frequently in the decay widths to be
discussed below.  The first function $f$ contains an extra factor in
the denominator to conform to the notation used in
Ref.~\cite{Ambrosanio:1997bq}.  The other functions are simply
integrals over $y$ with all relevant combinations of propagators in
the denominator and a numerator either constant or linear in $y$.


\section{\boldmath{$\Gamma \left(\tilde{\ell}_{\lowercase{i}}^- \to 
\tilde{\ell}_{\lowercase{j}}^- \ell_{\lowercase{k}}^-
\ell_{\lowercase{m}}^+ \right)$}}
\label{app:dileptonp}

\subsection{Total Width}

For the charge-preserving case, the matrix elements for all
contributing modes are presented in
Eqs.~(\ref{eqn:preservingMEstart})-(\ref{eqn:preservingMEend}).  The
total decay width is
\begin{eqnarray}
\Gamma \left(\tilde{\ell}_i^- \to \tilde{\ell}_j^- \ell_k^-
\ell_m^+\right) & = &\frac{1}{64 \pi^3 m_i} \intps |\mathcal{M}|^2 \\
& = &\frac{1}{64 \pi^3 m_i} \intps \left( \sum_{a,b=1}^4
\mathcal{M}_{\tilde{\chi}_a^0} \mathcal{M}_{\tilde{\chi}_b^0}^* +
\mathcal{M}_Z \mathcal{M}_Z^* +
\sum_{a,b} \mathcal{M}_{H_a} \mathcal{M}_{H_b}^* \right. \nonumber \\
&& \left.  + 2 \text{ Re} \left[\sum_{b=1}^4 \mathcal{M}_Z
\mathcal{M}_{\tilde{\chi}_b^0}^* + \sum_{a} \sum_{b=1}^4
\mathcal{M}_{H_a} \mathcal{M}_{\tilde{\chi}_b^0}^* + \sum_b
\mathcal{M}_Z \mathcal{M}_{H_b}^* \right] \right) \\ 
& = &\Gamma_{\tilde{\chi}^0 \tilde{\chi}^0} + \delta_{km}
\left(\Gamma_{ZZ} + \Gamma_{HH} + 2 \Gamma_{Z \tilde{\chi}^0} + 2
\Gamma_{H \tilde{\chi}^0} + 2 \Gamma_{ZH} \right) \ ,
\end{eqnarray}
where $\delta_{km}$ is the Kronecker delta function.  For
$\Gamma_{\tilde{\chi}^0 \tilde{\chi}^0}$ the general case of
independent $k$ and $m$ is taken; for the remaining widths, $k = m$ is
assumed to simplify the expressions.

\subsection{\boldmath{$\tilde{\chi}^0 \tilde{\chi}^0$} Contribution}

The neutralino width corresponds to the width given in
Ref.~\cite{Ambrosanio:1997bq} with the addition of flavor-violating
vertices.  The decay width is given by
\begin{equation}
\Gamma_{\tilde{\chi}^0 \tilde{\chi}^0} 
= \frac{m_{\tilde{\ell}_i}}{512 \pi^3} 
\sum_{t = 1}^6
\sum_{a, b = 1}^4 C_{\tilde{\chi}_a^0 \tilde{\chi}_b^0}^{(t)} 
I_{\tilde{\chi}_a^0 \tilde{\chi}_b^0}^{(t)}\, ,
\end{equation}
where $t$ labels the coefficients and integrals, and $a$ and $b$ label
the neutralinos.  The dimensionless integrals are defined by
\begin{eqnarray}
I_{\tilde{\chi}_a^0 \tilde{\chi}_b^0}^{(1)} &=& \int_{x_-}^{x_+} dx
\left(x - 2 r_k^2\right) \left(1 - x + r_k^2\right) \left(R^2 -
x\right) f \left(x, r_a, r_b\right) \\
I_{\tilde{\chi}_a^0 \tilde{\chi}_b^0}^{(2)} &=& r_a r_b
\int_{x_-}^{x_+} dx \left(x - 2 r_k^2\right) \left(R^2 - x\right)
f \left(x, r_a, r_b\right) \\ 
I_{\tilde{\chi}_a^0 \tilde{\chi}_b^0}^{(3)} &=& 2 r_m r_b
\int_{x_-}^{x_+} dx \left(x - 2 r_k^2\right) \left(1 - x +
r_k^2\right) f \left(x, r_a, r_b\right) \\ 
I_{\tilde{\chi}_a^0 \tilde{\chi}_b^0}^{(4)} &=& 2 r_k r_b
\int_{x_-}^{x_+} dx \left(1 - x + r_k^2\right) \left(R^2 - x\right)
f \left(x, r_a, r_b\right) \\ 
I_{\tilde{\chi}_a^0 \tilde{\chi}_b^0}^{(5)} &=& 2 r_k r_m r_a r_b
\int_{x_-}^{x_+} dx \left(1 - x + r_k^2\right) f \left(x, r_a,
r_b\right) \\ 
I_{\tilde{\chi}_a^0 \tilde{\chi}_b^0}^{(6)} &=& 2 r_k r_m
\int_{x_-}^{x_+} dx \left(1 - x + r_k^2\right)^2 f \left(x, r_a,
r_b\right)\, ,
\end{eqnarray}
with coefficients
\begin{eqnarray}
C_{\tilde{\chi}_a^0 \tilde{\chi}_b^0}^{(1)} &=& 
\beta_{aik}^{1*} \beta_{ajm}^1 \beta_{bik}^1 \beta_{bjm}^{1*} + 
\beta_{aik}^{2*} \beta_{ajm}^2 \beta_{bik}^2 \beta_{bjm}^{2*} \\
C_{\tilde{\chi}_a^0 \tilde{\chi}_b^0}^{(2)} &=& 
\beta_{aik}^{1*} \beta_{ajm}^2 \beta_{bik}^1 \beta_{bjm}^{2*} + 
\beta_{aik}^{2*} \beta_{ajm}^1 \beta_{bik}^2 \beta_{bjm}^{1*} \\
C_{\tilde{\chi}_a^0 \tilde{\chi}_b^0}^{(3)} &=& 2 \text{ Re} \left[
\beta_{aik}^{1*} \beta_{ajm}^1 \beta_{bik}^1 \beta_{bjm}^{2*} + 
\beta_{aik}^{2*} \beta_{ajm}^2 \beta_{bik}^2 \beta_{bjm}^{1*} \right] \\
C_{\tilde{\chi}_a^0 \tilde{\chi}_b^0}^{(4)} &=& - 2 \text{ Re} \left[
\beta_{aik}^{1*} \beta_{ajm}^1 \beta_{bik}^2 \beta_{bjm}^{1*} + 
\beta_{aik}^{2*} \beta_{ajm}^2 \beta_{bik}^1 \beta_{bjm}^{2*} \right] \\
C_{\tilde{\chi}_a^0 \tilde{\chi}_b^0}^{(5)} &=& - 4 \text{ Re} \left[
\beta_{aik}^{1*} \beta_{ajm}^2 \beta_{bik}^2 \beta_{bjm}^{1*} \right] \\
C_{\tilde{\chi}_a^0 \tilde{\chi}_b^0}^{(6)} &=& - 4 \text{ Re} \left[
\beta_{aik}^{1*} \beta_{ajm}^1 \beta_{bik}^2 \beta_{bjm}^{2*} \right] \ .
\end{eqnarray}

\subsection{\boldmath{$ZZ$} Contribution}

The $Z$ boson contribution to the decay width is
\begin{equation}
\Gamma_{ZZ} = \frac{m_i}{512 \pi^3} 
\left| \frac{g \zeta_{ji}^{(2)}}{\cos \theta_W} \right|^2 
\sum_{t = 1}^2 C_{ZZ}^{(t)} I_{ZZ}^{(t)} \ ,
\end{equation}
where the dimensionless integrals are
\begin{eqnarray}
I_{ZZ}^{(1)} & = &\int_{x_-}^{x_+} dx \, 
\left(2x - 2 - r_k^2\right) i^{(2)} \left(x,r_Z\right) \nonumber \\
&& + \left[ 2 - \left(2 + r_k^2\right) x - 2 r_j^2 + 3 r_k^2 + r_k^2
r_j^2 \right] i^{(1)} \left(x,r_Z\right) \\
I_{ZZ}^{(2)} & = &r_k^2 \int_{x_-}^{x_+} dx \, i^{(2)}
\left(x,r_Z\right) + \left(x - 3 - r_j^2\right)
i^{(1)}\left(x,r_Z\right) \, ,
\end{eqnarray}
and the coefficients are
\begin{eqnarray}
C_{ZZ}^{(1)} & = &8 \sin^4 \theta_W - 4 \sin^2 \theta_W + 1 \\
C_{ZZ}^{(2)} & = &8 \sin^4 \theta_W - 4 \sin^2 \theta_W \ .
\end{eqnarray}

\subsection{\boldmath{$HH$} Contribution}

The purely Higgs-mediated contribution actually consists of several
pieces with similar phase space structure.  Furthermore, it simplifies
into the sum of contributions from the real and pseudoscalar Higgs
bosons, since the interference term between a real scalar and a
pseudoscalar vanishes.

The width is given by
\begin{eqnarray}
\Gamma_{HH} = \frac{1}{256 \pi^3 m_i} \left( C_{h^0 h^0} I_{h^0
h^0}^r + C_{H^0 H^0} I_{H^0}^r + 2 C_{h^0 H^0} I_{h^0 H^0}^r
\right. \nonumber \\
 + \left. C_{A^0 A^0} I_{A^0 A^0}^p + C_{G^0 G^0} I_{G^0 G^0}^p + 2
C_{A^0 G^0} I_{A^0 G^0}^p \right),
\end{eqnarray}
where the dimensionless integrals are
\begin{eqnarray}
I_{H_a H_a}^r & = &\int_{x_-}^{x_+} dx \, i^{(2)}
\left(x,r_{H_a}\right) + \left(x - R^2 - 2 r_k^2\right) i^{(1)}
\left(x,r_{H_a}\right) \\
I_{H_a H_b}^r & = &\int_{x_-}^{x_+} dx \, h^{(2)}
\left(x,r_{H_a},r_{H_b}\right) + \left(x - R^2 - 2 r_k^2\right)
h^{(1)} \left(x,r_{H_a},r_{H_b}\right) \\ 
I_{H_a H_a}^p & = &\int_{x_-}^{x_+} dx \, i^{(2)}
\left(x,r_{H_a}\right) + \left(x - 1 + r_j^2\right) i^{(1)}
\left(x,r_{H_a}\right) \\ 
I_{H_a H_b}^p & = &\int_{x_-}^{x_+} dx \, h^{(2)}
\left(x,r_{H_a},r_{H_b}\right) + \left(x - 1 + r_j^2\right) h^{(1)}
\left(x,r_{H_a},r_{H_b}\right) \ ,
\end{eqnarray}
where $a \neq b$, and the coefficients are
\begin{eqnarray}
C_{h^0 h^0} & =& \left| \sigma_{ji}^{(2)} y_k^{(\ell)} \sin \theta_H
\right|^2 \\ 
C_{H^0 H^0} & =& \left| \sigma_{ji}^{(3)} y_k^{(\ell)} \cos \theta_H
\right|^2 \\ 
C_{h^0 H^0} & =&- \text{Re}\left[ \sigma_{ji}^{(2)}
\sigma_{ji}^{(3)*} y_k^{(\ell)2} \sin \theta_H \cos \theta_H\right] \\
C_{A^0 A^0} & =& \left| \sigma_{ji}^{(4)} y_k^{(\ell)} \sin \beta
\right|^2 \\ 
C_{G^0 G^0} & =& \left| \sigma_{ji}^{(5)} y_k^{(\ell)} \cos \beta
\right|^2 \\ 
C_{A^0 G^0} & = &- \text{Re}\left[ \sigma_{ji}^{(4)}
\sigma_{ji}^{(5)*} y_k^{(\ell)2} \sin \beta \cos \beta\right] \ .
\end{eqnarray}

\subsection{\boldmath{$Z \tilde{\chi}^0$} Contribution}

The width from the $Z \tilde{\chi}^0$ interference term is
\begin{equation}
\Gamma_{Z \tilde{\chi}_b^0} = - \frac{m_i}
{512 \pi^3} 
\frac{g \zeta_{ji}^{(2)}}{\cos \theta_W} 
\sum_{b = 1}^4 
\sum_{t = 1}^2 C_{Z \tilde{\chi}_b^0}^{(t)} 
I_{Z \tilde{\chi}_b^0}^{(t)}\, ,
\end{equation}
where the integrals are
\begin{eqnarray}
\nonumber I_{Z \tilde{\chi}_b^0}^{(1)} & = &\int_{x_-}^{x_+} dx \,
\left[ 2 \left( 1 - x + r_k^2\right) g^{(2)} \left(x,r_Z,r_b\right) 
\right. \\
& &  \left. 
+ \left(2 x - 2 R^2 + r_k^2 - r_j^2 r_k^2 \right)
g^{(1)}\left(x,r_Z,r_b\right) \right]\\
I_{Z \tilde{\chi}_b^0}^{(2)} & = &r_k r_b \int_{x_-}^{x_+} dx \,
\left[ g^{(2)} \left(x,r_Z,r_b\right) - \left(1 + x - r_j^2\right)
g^{(1)}\left(x,r_Z,r_b\right) \right] \\
I_{Z \tilde{\chi}_b^0}^{(3)} & = &r_k r_b \int_{x_-}^{x_+} dx \,
\left[ g^{(2)}\left(x,r_Z,r_b\right) + \left(1 - x - r_j^2\right)
g^{(1)} \left(x,r_Z,r_b\right) \right] \\ 
I_{Z \tilde{\chi}_b^0}^{(4)} & = &r_k^2 \int_{x_-}^{x_+} dx \, \left[
\left(3 - 2x + r_j^2\right) g^{(1)} \left(x,r_Z,r_b\right) \right] \ ,
\end{eqnarray}
and the coefficients are
\begin{eqnarray}
C_{Z \tilde{\chi}_b^0}^{(1)} & = &\text{ Re} \left[\left(2 \sin^2
\theta_W - 1\right) \beta_{bik}^1 \beta_{bjm}^{1*} + 2 \sin^2 \theta_W
\beta_{bik}^2 \beta_{bjm}^{2*}\right] \\
C_{Z \tilde{\chi}_b^0}^{(2)} & = &\text{ Re} \left[\left(2 \sin^2
\theta_W - 1\right) \beta_{bik}^1 \beta_{bjm}^{2*} + 2 \sin^2 \theta_W
\beta_{bik}^2 \beta_{bjm}^{1*}\right] \\ 
C_{Z \tilde{\chi}_b^0}^{(3)} & = &\text{ Re} \left[2 \sin^2 \theta_W
\beta_{bik}^1 \beta_{bjm}^{2*} + \left(2 \sin^2 \theta_W - 1\right)
\beta_{bik}^2 \beta_{bjm}^{1*}\right] \\ 
C_{Z \tilde{\chi}_b^0}^{(4)} & = &\text{ Re} \left[2 \sin^2 \theta_W
\beta_{bik}^1 \beta_{bjm}^{1*} + \left(2 \sin^2 \theta_W - 1\right)
\beta_{bik}^2 \beta_{bjm}^{2*}\right] \ .
\end{eqnarray}

\subsection{\boldmath{$H \tilde{\chi}^0$} Contribution}
The width from the $H \tilde{\chi}^0$ interference is
\begin{equation}
\Gamma_{H \tilde{\chi}^0} = \frac{1}{
256 \sqrt{2} \pi^3} \sum _{b = 1}^4
\sum _{t = 1}^2 \left(
C_{h^0 \tilde{\chi}_b^0}^{(t)} I_{h^0 \tilde{\chi}_b^0}^{H(t)}
+ C_{H^0 \tilde{\chi}_b^0}^{(t)} I_{H^0 \tilde{\chi}_b^0}^{H(t)}
+ C_{A^0 \tilde{\chi}_b^0}^{(t)} I_{A^0 \tilde{\chi}_b^0}^{P(t)}
+ C_{G^0 \tilde{\chi}_b^0}^{(t)} I_{G^0 \tilde{\chi}_b^0}^{P(t)}
\right) \ ,
\end{equation}
where the dimensionless integrals are
\begin{eqnarray}
I_{H_a \tilde{\chi}_b^0}^{H(1)} & =& r_b \int_{x_-}^{x_+} dx \,
\left[ g^{(2)} \left(x,r_{H_a},r_b\right) + \left(x - R^2 - 2
r_k^2\right) g^{(1)} \left(x,r_{H_a},r_b\right) \right] \\
I_{H_a \tilde{\chi}_b^0}^{H(2)} & =& r_k \int_{x_-}^{x_+} dx \,
\left(2 x - R^2 - 2 r_k^2\right) g^{(1)} \left(x,r_{H_a},r_b\right) \\
I_{H_a \tilde{\chi}_b^0}^{P(1)} & =& r_b \int_{x_-}^{x_+} dx \,
\left[ g^{(2)} \left(x,r_{H_a},r_b\right) + \left(x - 1 + r_j^2\right)
g^{(1)} \left(x,r_{H_a},r_b\right) \right] \\ 
I_{H_a \tilde{\chi}_b^0}^{P(2)} & =& r_k \int_{x_-}^{x_+} dx \,
\left(1 - r_j^2\right) g^{(1)} \left(x,r_{H_a},r_b\right)\, ,
\end{eqnarray}
and the coefficients are
\begin{eqnarray}
C_{h^0 \tilde{\chi}_b^0}^{(1)} & = & \text{ Re}\left[
\sigma_{ji}^{(2)} y_k^{(\ell)} \sin \theta_H \left(\beta_{bik}^1
\beta_{bjm}^{2*} + \beta_{bik}^2 \beta_{bjm}^{1*}\right)\right] \\
C_{h^0 \tilde{\chi}_b^0}^{(2)} & = & \text{ Re}\left[
\sigma_{ji}^{(2)} y_k^{(\ell)} \sin \theta_H \left(\beta_{bik}^1
\beta_{bjm}^{1*} + \beta_{bik}^2 \beta_{bjm}^{2*}\right)\right] \\
C_{H^0 \tilde{\chi}_b^0}^{(1)} & = & - \text{Re}\left[
\sigma_{ji}^{(3)} y_k^{(\ell)} \cos \theta_H \left(\beta_{bik}^1
\beta_{bjm}^{2*} + \beta_{bik}^2 \beta_{bjm}^{1*}\right)\right] \\
C_{H^0 \tilde{\chi}_b^0}^{(2)} & = & - \text{Re}\left[
\sigma_{ji}^{(3)} y_k^{(\ell)} \cos \theta_H \left(\beta_{bik}^1
\beta_{bjm}^{1*} + \beta_{bik}^2 \beta_{bjm}^{2*}\right)\right] \\
C_{A^0 \tilde{\chi}_b^0}^{(1)} & = & \text{ Re}\left[
\sigma_{ji}^{(4)} y_k^{(\ell)} \sin \beta \left(\beta_{bik}^1
\beta_{bjm}^{2*} - \beta_{bik}^2 \beta_{bjm}^{1*}\right)\right] \\
C_{A^0 \tilde{\chi}_b^0}^{(2)} & = & \text{ Re}\left[
\sigma_{ji}^{(4)} y_k^{(\ell)} \sin \beta \left(\beta_{bik}^1
\beta_{bjm}^{1*} - \beta_{bik}^2 \beta_{bjm}^{2*}\right)\right] \\
C_{G^0 \tilde{\chi}_b^0}^{(1)} & = & - \text{Re}\left[
\sigma_{ji}^{(5)} y_k^{(\ell)} \cos \beta \left(\beta_{bik}^1
\beta_{bjm}^{2*} - \beta_{bik}^2 \beta_{bjm}^{1*}\right)\right] \\
C_{G^0 \tilde{\chi}_b^0}^{(2)} & = & - \text{Re}\left[
\sigma_{ji}^{(5)} y_k^{(\ell)} \cos \beta \left(\beta_{bik}^1
\beta_{bjm}^{1*} - \beta_{bik}^2 \beta_{bjm}^{2*}\right)\right]\ .
\end{eqnarray}

\subsection{\boldmath{$ZH$} Contribution}
The $Z H$ interference decay width is
\begin{equation}
\Gamma_{Z H} = \frac{g \zeta_{ji}^{(2)}} {256 \sqrt{2} \pi^3 \cos
\theta_W} \left( C_{Z h^0} I_{Z h^0}^r + C_{Z H^0} I_{Z H^0}^r + C_{Z
A^0} I_{Z A^0}^p + C_{Z G^0} I_{Z G^0}^p \right) \, ,
\end{equation}
where
\begin{eqnarray}
I_{Z H_a}^r & =& r_k \int_{x_-}^{x_+} dx \, \left[ h^{(2)}
\left(x,r_Z,r_{H_a}\right) - x h^{(1)} \left(x,r_Z,r_{H_a}\right) 
\right] \\
I_{Z H_a}^p & =& r_k \int_{x_-}^{x_+} dx \, \left(1 - r_j^2\right)
h^{(1)} \left(x,r_Z,r_{H_a}\right) \, ,
\end{eqnarray}
and 
\begin{eqnarray}
C_{Z h^0} & =& - \text{Re}\left[\sigma_{ji}^{(2)} y_k^{(\ell)} \sin
\theta_H \left(4 \sin^2 \theta_W - 1\right)\right] \\ 
C_{Z H^0} & = & \text{ Re}\left[\sigma_{ji}^{(3)} y_k^{(\ell)} \cos
\theta_H \left(4 \sin^2 \theta_W - 1\right)\right] \\ 
C_{Z A^0} & = &- \text{Re}\left[\sigma_{ji}^{(4)} y_k^{(\ell)} \sin
\beta\right] \\ 
C_{Z G^0} & = & \text{ Re}\left[\sigma_{ji}^{(5)} y_k^{(\ell)} \cos
\beta\right] \ .
\end{eqnarray}


\section{\boldmath{$\Gamma \left(\tilde{\ell}_{\lowercase{i}}^- \to 
\tilde{\ell}_{\lowercase{j}}^+ \ell_{\lowercase{k}}^-
\ell_{\lowercase{m}}^-\right)$}}
\label{app:dileptonf}

\subsection{Total Width}

For the charge-flipping decay, the calculation is complicated by the
fact that the same-sign daughter leptons create an interference term.
It is convenient to break the matrix element into two parts, as shown
in Eqs.~(\ref{eqn:flippingMEstart})-(\ref{eqn:flippingMEend}).  It is
correspondingly convenient to separate the decay width into three
terms
\begin{equation} 
\Gamma \left(\tilde{\ell}_i^- \to
\tilde{\ell}_j^+ \ell_k^- \ell_m^-\right) = C_{km} \left( \Gamma_{11}
+ \Gamma_{22} - 2 \Gamma_{21} \right) \ ,
\end{equation}
where
\begin{equation}
\Gamma_{ij} = \frac{1}{64 \pi^3 m_i} \intps \sum_{a,b=1}^4 \text{ Re}
\left[\mathcal{M}_a^{(i)} \mathcal{M}_b^{(j)*} \right] \ ,
\end{equation}
and $C_{km}$ is 1 when the two outgoing leptons are of different
generations and $1/2$ when they are of the same generation.

\subsection{\boldmath{$\Gamma_{11}$} Width}

{}From inspection of the matrix element, $\mathcal{M}_a^{(i)}$ is
identical to $\mathcal{M}_{\tilde{\chi}_a^0}$ except for the
coefficients in the $ajm$ vertex.  Then $\Gamma_{11}$ is identical to
$\Gamma_{\tilde{\chi}^0 \tilde{\chi}^0}$ in the charge-preserving case
with the substitutions
\begin{eqnarray}
\beta_{xjm}^1 \to \beta_{xjm}^{2 *}\, , \quad \beta_{xjm}^2 \to
\beta_{xjm}^{1*}\, ,
\end{eqnarray}
and with the identical substitutions for the complex conjugates, where
$x = a , b$.

\subsection{\boldmath{$\Gamma_{22}$} Width}

{}From further inspection of the matrix elements, $\mathcal{M}_a^{(1)}$
and $\mathcal{M}_a^{(2)}$ differ only in the interchange of the two
outgoing leptons.  Thus $\Gamma_{22}$ may be obtained from
$\Gamma_{11}$ with the interchange $k \leftrightarrow m$.

\subsection{\boldmath{$\Gamma_{21}$} Width}
\label{app:gamma21}

The $\Gamma_{21}$ width from the interference term is given by
\begin{eqnarray}
\Gamma_{21} & = &\frac{1}{64 \pi^3 m_i} \intps \sum_{a,b} \text{ Re}
\left[ \mathcal{M}_a^{(2)}\mathcal{M}_b^{(1)*} \right] \nonumber \\ 
& = & \frac{m_i}{256 \pi^3} \sum_{t=1}^8 \sum_{a,b=1}^4 \text{ Re}
\left[ D_{\tilde{\chi}_a^0 \tilde{\chi}_b^0}^{(t)} J_{\tilde{\chi}_a^0
\tilde{\chi}_b^0}^{(t)} \right] \ ,
\end{eqnarray}
where the integrals are
\begin{eqnarray}
\nonumber J_{\tilde{\chi}_a^0 \tilde{\chi}_b^0}^{(1)} & = &
\int_{x_-}^{x_+} dx \, \left\{ 
\left[ 2 r_k^2 r_m^2 - x \left( 1 + r_m^2
\right) + R^2\right] f^{(1)} \left(x, r_a,r_b\right) \right\} \\
& & - \left(1 - x + r_k^2\right) f^{(2)} \left(x, r_a,r_b\right) \\
J_{\tilde{\chi}_a^0 \tilde{\chi}_b^0}^{(2)} & = & r_k r_m
\int_{x_-}^{x_+} dx \, \left(1 + r_j^2 - r_k^2 - r_m^2 \right) f^{(1)}
\left(x, r_a,r_b\right) \\
J_{\tilde{\chi}_a^0 \tilde{\chi}_b^0}^{(3)} & = & r_m r_b
\int_{x_-}^{x_+} dx \, \left[
R^2 f^{(1)} \left(x, r_a,r_b\right) - f^{(2)}
\left(x, r_a,r_b\right) \right] \\
J_{\tilde{\chi}_a^0 \tilde{\chi}_b^0}^{(4)} & = & r_k r_b
\int_{x_-}^{x_+} dx \, \left[ f^{(2)} \left(x, r_a,r_b\right) - 2 r_m^2
f^{(1)} \left(x, r_a,r_b\right) \right] \\
J_{\tilde{\chi}_a^0 \tilde{\chi}_b^0}^{(5)} & = & r_m r_a
\int_{x_-}^{x_+} dx \, \left(x - 2 r_k^2\right) f^{(1)} \left(x,
r_a,r_b\right) \\
J_{\tilde{\chi}_a^0 \tilde{\chi}_b^0}^{(6)} & = & r_k r_a
\int_{x_-}^{x_+} dx \, \left(R^2 - x\right) f^{(1)} \left(x,
r_a,r_b\right) \\
J_{\tilde{\chi}_a^0 \tilde{\chi}_b^0}^{(7)} & = & r_a r_b
\int_{x_-}^{x_+} dx \, \left[ \left(x - R^2\right) f^{(1)} \left(x,
r_a,r_b\right) + f^{(2)} \left(x, r_a,r_b\right) \right] \\
J_{\tilde{\chi}_a^0 \tilde{\chi}_b^0}^{(8)} & = & 2 r_k r_m r_a r_b
\int_{x_-}^{x_+} dx \, f^{(1)} \left(x, r_a,r_b\right) \, ,
\end{eqnarray}
and the coefficients are
\begin{eqnarray}
D_{\tilde{\chi}_a^0 \tilde{\chi}_b^0}^{(1)} 
&=& \beta_{aim}^{1*} \beta_{ajk}^{2*} \beta_{bik}^2 \beta_{bjm}^1 
+ \beta_{aim}^{2*} \beta_{ajk}^{1*} \beta_{bik}^1 \beta_{bjm}^2 \\ 
D_{\tilde{\chi}_a^0 \tilde{\chi}_b^0}^{(2)} 
&=& - \left( \beta_{aim}^{1*} \beta_{ajk}^{2*} 
\beta_{bik}^1 \beta_{bjm}^2 + \beta_{aim}^{2*} \beta_{ajk}^{1*} 
\beta_{bik}^2 \beta_{bjm}^1 \right) \\
D_{\tilde{\chi}_a^0 \tilde{\chi}_b^0}^{(3)} 
&=& \beta_{aim}^{1*} \beta_{ajk}^{2*} \beta_{bik}^2 \beta_{bjm}^2 
+ \beta_{aim}^{2*} \beta_{ajk}^{1*} \beta_{bik}^1 \beta_{bjm}^1 \\
D_{\tilde{\chi}_a^0 \tilde{\chi}_b^0}^{(4)} 
&=& - \left( \beta_{aim}^{1*} \beta_{ajk}^{2*} 
\beta_{bik}^1 \beta_{bjm}^1 + \beta_{aim}^{2*} \beta_{ajk}^{1*} 
\beta_{bik}^2 \beta_{bjm}^2 \right) \\
D_{\tilde{\chi}_a^0 \tilde{\chi}_b^0}^{(5)} 
&=& - \left( \beta_{aim}^{1*} \beta_{ajk}^{1*} 
\beta_{bik}^1 \beta_{bjm}^2 + \beta_{aim}^{2*} \beta_{ajk}^{2*} 
\beta_{bik}^2 \beta_{bjm}^1 \right) \\
D_{\tilde{\chi}_a^0 \tilde{\chi}_b^0}^{(6)} 
&=& \beta_{aim}^{1*} \beta_{ajk}^{1*} \beta_{bik}^2 \beta_{bjm}^1 
+ \beta_{aim}^{2*} \beta_{ajk}^{2*} \beta_{bik}^1 \beta_{bjm}^2 \\
D_{\tilde{\chi}_a^0 \tilde{\chi}_b^0}^{(7)} 
&=& - \left( \beta_{aim}^{1*} \beta_{ajk}^{1*} 
\beta_{bik}^1 \beta_{bjm}^1 + \beta_{aim}^{2*} \beta_{ajk}^{2*} 
\beta_{bik}^2 \beta_{bjm}^2 \right) \\
D_{\tilde{\chi}_a^0 \tilde{\chi}_b^0}^{(8)} 
&=& \beta_{aim}^{1*} \beta_{ajk}^{1*} \beta_{bik}^2 \beta_{bjm}^2 
+ \beta_{aim}^{2*} \beta_{ajk}^{2*} \beta_{bik}^1 \beta_{bjm}^1 \ .
\end{eqnarray}


\section{\boldmath{$\Gamma\left(\tilde{\ell}_{\lowercase{i}}^- \to 
\tilde{\ell}_{\lowercase{j}}^- \nu_{\lowercase{k}} \,
\overline{\nu}_{\lowercase{m}} \right)$}}
\label{app:nunubar}

\subsection{Matrix Elements}

The decay to neutrinos is mediated by charginos $\tilde{\chi}_a^-$ and
the $Z$ boson, and so the matrix element is
\begin{equation}
\mathcal{M} = \sum_{a=1}^2 \mathcal{M}_{\tilde{\chi}_a^-} 
+ \mathcal{M}_Z \ ,
\end{equation}
where
\begin{eqnarray}
\mathcal{M}_{\tilde{\chi}_a^-} & = & - i \overline{u} \left(p_k\right)
\left(i \gamma_{aik}^* P_R \right) \frac{\left( \slashed{p}_k -
\slashed{p}_i \right) + m_a}{\left(p_i - p_k\right)^2 - m_a^2} \left(i
\gamma_{ajm} P_L \right) v\left(p_m\right) \nonumber \\
& = & i \left(\gamma_{aik}^* \gamma_{ajm}\right) \overline{u}
\left(p_k\right) \frac{\slashed{p}_k - \slashed{p}_i} {\left(p_i -
p_k\right)^2 - m_a^2} P_L v\left(p_m\right) \\ 
\mathcal{M}_Z & = & \frac{i g \zeta_{ji}^{(2)}}{ 2 \cos \theta_W}
\frac{ 1} {\left(p_i - p_j\right)^2 - m_Z^2} \overline{u}
\left(p_k\right) \left(\slashed{p}_i + \slashed{p}_j\right) P_L v
\left(p_m\right) \ .
\end{eqnarray}

\subsection{Total Width}

The width is therefore
\begin{eqnarray}
\lefteqn{ \Gamma \left(\tilde{\ell}_i^- \to \tilde{\ell}_j^- \nu_k
\overline{\nu}_m\right) = \frac{1}{64 \pi^3 m_i} \intps \left|
\mathcal{M} \right|^2} \nonumber \\
\quad & = & \frac{1}{64 \pi^3 m_i} \intps \left( \sum_{a,b=1}^4
\mathcal{M}_{\tilde{\chi}_a^-} \mathcal{M}_{\tilde{\chi}_b^-}^* +
\mathcal{M}_Z \mathcal{M}_Z^* + 2 \sum_{b=1}^4 \text{
Re}\left[\mathcal{M}_Z \mathcal{M}_{\tilde{\chi}_b^-}^*\right] \right)
\nonumber \\
& = & \Gamma_{\tilde{\chi}^- \tilde{\chi}^-} + \delta_{km} \left( 
\Gamma_{ZZ} + 2 \Gamma_{Z \tilde{\chi}^-} \right) \ ,
\end{eqnarray}
where the partial widths are defined below.

\subsection{\boldmath{$\tilde{\chi}^- \tilde{\chi}^-$} Contribution}

The chargino-mediated width is
\begin{eqnarray}
\Gamma_{\tilde{\chi}^- \tilde{\chi}^-} & = & \frac{m_i}{512 \pi^3}
\sum_{a,b=1}^2 \gamma_{aik}^* \gamma_{ajm} \gamma_{bik} \gamma_{bjm}^*
\int_0^{1 - r_j^2} dx \, \frac{ x^2 \left( 1 - x - r_j^2 \right)^2}
{\left( 1 - x \right)\left( 1 - x - r_a^2 \right)\left(1 - x - r_b^2
\right)} \nonumber \\
& = & \frac{m_i}{512 \pi^3} \sum_{a,b=1}^2 \gamma_{aik}^* \gamma_{ajm}
\gamma_{bik} \gamma_{bjm}^* \int_{x_-}^{x_+} dx \, x \left(1 -
x\right) \left(1 - x - r_j^2\right) f \left(x,r_a,r_b\right) \nonumber
\\
& = & \frac{m_i}{512 \pi^3} \sum_{a,b=1}^2 \gamma_{aik}^* \gamma_{ajm}
\gamma_{bik} \gamma_{bjm}^* I_{\tilde{\chi}_a^- \tilde{\chi}_b^-} \ .
\end{eqnarray}
Here the integral $I_{{\tilde{\chi}_a^-} {\tilde{\chi}_b^-}}$ is the
same as $I_{{\tilde{\chi}_a^0} {\tilde{\chi}^0_b}}^{(1)}$ for the
charge-preserving case, except the neutralino mass is replaced by the
chargino mass and lepton masses $m_k$ and $m_m$ are set to zero.

\subsection{\boldmath{$ZZ$} Contribution}

The $Z$-mediated width is
\begin{eqnarray}
\Gamma_{ZZ} & = &\frac{m_i}{256 \pi^3} 
\int_{x_-}^{x_+} 
dx \, \int_{y_-}^{y_+} dy \, \mathcal{M}_Z \mathcal{M}_Z^* \nonumber \\
& = &\frac{m_i}{512 \pi^3} 
\left| \frac{g \zeta_{ji}^{(2)}}
{\cos \theta_W} \right|^2
\int_{x_-}^{x_+} dx \, \left[ 2 \left(x - 1\right) i^{(2)} \left(x,r_Z\right) 
+ 2 \left(1 - x - r_j^2\right) i^{(1)} \left(x,r_Z\right) \right] \nonumber \\
& = &\frac{m_i}{512 \pi^3} 
\left| \frac{g \zeta_{ji}^{(2)}}
{\cos \theta_W} \right|^2 I_{ZZ}^{(1)} \ .
\end{eqnarray}
Here again the integral is the same as the $ZZ$ integral for the
charge-preserving case except the lepton masses $m_k$ and $m_m$ are
set to zero.

\subsection{\boldmath{$Z \tilde{\chi}^-$} Contribution}

The $Z$-chargino interference term is 
\begin{eqnarray}
\Gamma_{Z \tilde{\chi}_b^-} & = &\frac{m_i}{ 256 \pi^3}
\int_{x_-}^{x_+} dx \, \int_{y_-}^{y_+} dy \, \mathcal{M}_Z
\mathcal{M}_{\tilde{\chi}_b^-}^* \nonumber \\
& = &- \frac{m_i}{512 \pi^3} \frac{g \zeta_{ji}^{(2)}}{\cos \theta_W}
\text{ Re}\left[ \gamma_{bik} \gamma_{bjm}^*\right] \nonumber \\
& & \int_{x_-}^{x_+} dx \, \left[ 2 \left(1 - x\right) g^{(2)}
\left(x,r_Z,r_b\right) + 2 \left(x - 1 + r_j^2\right)
g^{(1)}\left(x,r_Z,r_b\right) \right] \nonumber \\
& = &- \frac{m_i}{512 \pi^3} \text{ Re} \left[ \frac{g
\zeta_{ji}^{(2)}}{ \cos \theta_W} \gamma_{bik} \gamma_{bjm}^* \right]
I_{Z \tilde{\chi}_b^-} \ .
\end{eqnarray}
Here the integral $I_{Z \tilde{\chi}_b^-}$ is the same as $I_{Z
\tilde{\chi}_b^0}^{(1)}$ for the charge-preserving case, except the
neutralino mass is replaced by the chargino mass and lepton masses
$m_k$ and $m_m$ are set to zero.


\section{\boldmath{$\Gamma\left(\tilde{\ell}_{\lowercase{i}}^- \to 
\tilde{\ell}_{\lowercase{j}}^- \lowercase{q}_{\lowercase{k}} \,
\overline{\lowercase{q}}_{\lowercase{m}} \right)$}}
\label{app:qqbar}

\subsection{Matrix Elements}

The decay modes with daughter quarks are much the same as those for
daughter leptons except the neutralino intermediary contribution is
removed and the different couplings are substituted.  The matrix
elements for the up-type quarks are

\begin{eqnarray}
\mathcal{M}_Z &=& \frac{3 i g \zeta_{ji}^{(2)}}{2 \cos \theta_W}
\frac{1}{\left(p_i - p_j\right)^2 - m_Z^2} \overline{u}
\left(p_k\right) \left(\slashed{p}_i + \slashed{p}_j\right) \left(
P_L - \frac{4}{3} \sin^2 \theta_W \right) v \left(p_m\right)
\delta_{km} \\
\mathcal{M}_{h^0} &=& \frac{-3 i \sigma_{ji}^{(2)} y_k^{(u)} \cos
\theta_H}{\sqrt{2}} \frac{1}{\left(p_i - p_j\right)^2 - m_{h^0}^2}
\overline{u} \left(p_k\right) v \left(p_m\right) \delta_{km} \\
\mathcal{M}_{H^0} &=& \frac{-3 i \sigma_{ji}^{(3)} y_k^{(u)} \sin
\theta_H}{\sqrt{2}} \frac{1}{\left(p_i - p_j\right)^2 - m_{H^0}^2}
\overline{u} \left(p_k\right) v \left(p_m\right) \delta_{km} \\
\mathcal{M}_{A^0} &=& \frac{-3 i \sigma_{ji}^{(4)} y_k^{(u)} \cos
\beta}{\sqrt{2}} \frac{1}{\left(p_i - p_j\right)^2 - m_{A^0}^2}
\overline{u} \left(p_k\right) \gamma^5 v \left(p_m\right) \delta_{km}
\\
\mathcal{M}_{G^0} &=& \frac{-3 i \sigma_{ji}^{(5)} y_k^{(u)} \sin
\beta}{\sqrt{2}} \frac{1}{\left(p_i - p_j\right)^2 - m_Z^2}
\overline{u} \left(p_k\right) \gamma^5 v \left(p_m\right) \delta_{km}
\ ,
\end{eqnarray}
where the factor of 3 is the color factor, and those for the down-type
quarks are

\begin{eqnarray}
\mathcal{M}_Z &=& \frac{3 i g \zeta_{ji}^{(2)}}{2 \cos \theta_W}
\frac{1}{\left(p_i - p_j\right)^2 - m_Z^2} \overline{u}
\left(p_k\right) \left(\slashed{p}_i + \slashed{p}_j\right) \left(
\frac{2}{3} \sin^2 \theta_W - P_L \right) v \left(p_m\right)
\delta_{km} \\
\mathcal{M}_{h^0} &=& \frac{3 i \sigma_{ji}^{(2)} y_k^{(d)} \sin
\theta_H}{\sqrt{2}} \frac{1}{\left(p_i - p_j\right)^2 - m_{h^0}^2}
\overline{u} \left(p_k\right) v \left(p_m\right) \delta_{km} \\
\mathcal{M}_{H^0} &=& \frac{-3 i \sigma_{ji}^{(3)} y_k^{(d)} \cos
\theta_H}{\sqrt{2}} \frac{1}{\left(p_i - p_j\right)^2 - m_{H^0}^2}
\overline{u} \left(p_k\right) v \left(p_m\right) \delta_{km} \\
\mathcal{M}_{A^0} &=& \frac{3i \sigma_{ji}^{(4)} y_k^{(d)} \sin
\beta}{\sqrt{2}} \frac{1}{\left(p_i - p_j\right)^2 - m_{A^0}^2}
\overline{u} \left(p_k\right) \gamma^5 v \left(p_m\right) \delta_{km}
\\
\mathcal{M}_{G^0} &=& \frac{-3 i \sigma_{ji}^{(5)} y_k^{(d)} \cos
\beta}{\sqrt{2}} \frac{1}{\left(p_i - p_j\right)^2 - m_Z^2}
\overline{u} \left(p_k\right) \gamma^5 v \left(p_m\right) \delta_{km}
\ .
\end{eqnarray}

The total decay width is then
\begin{equation}
\Gamma \left( \tilde{\ell}_i^- \to \tilde{\ell}_j^- q_k \overline{q}_k
\right) = \Gamma_{ZZ} + \Gamma_{HH} + 2 \Gamma_{ZH}.
\end{equation}

\subsection{Up-Type Quarks}

The decay width to up-type quarks is
\begin{eqnarray}
\lefteqn{\Gamma \left( \tilde{\ell}_i^- \to \tilde{\ell}_j^- u_k
\overline{u}_k \right) = \frac{9 m_i}{512 \pi^3} \left| \frac{g
\zeta_{ji}^{(2)}}{ \cos \theta_W} \right|^2 \sum_{t = 1}^2
A_{ZZ}^{(t)} I_{ZZ}^{(t)} } \nonumber \\ 
&& + \frac{9}{256 \pi^3 m_i} \left( A_{h^0 h^0} I_{h^0 h^0}^r + A_{H^0
H^0} I_{H^0 H^0}^r + 2 A_{h^0 H^0} I_{h^0 H^0}^r \right. \nonumber \\
&& \left. \qquad + A_{A^0 A^0} I_{A^0 A^0}^p + A_{G^0 G^0} I_{G^0
G^0}^p + 2 A_{A^0 G^0} I_{A^0 G^0}^p \right) \nonumber \\
&& + 2 \left[ \frac{9 g \zeta_{ji}^{(2)}}{ 256 \sqrt{2} \pi^3 \cos
\theta_W} \left( A_{Z h^0} I_{Z h^0}^r + A_{Z H^0} I_{Z H^0}^r + A_{Z
A^0} I_{Z A^0}^p + A_{Z G^0} I_{Z G^0}^p \right) \right] \ ,
\end{eqnarray}
with integrals from \appref{dileptonp} with quark masses substituted.
The coefficients are
\begin{eqnarray}
A_{ZZ}^{(1)} & = & \frac{32}{9} \sin^4 \theta_W - \frac{8}{3} \sin^2
\theta_W + 1 \\
A_{ZZ}^{(2)} & = & \frac{32}{9} \sin^4 \theta_W - \frac{8}{3} \sin^2
\theta_W \\
A_{h^0 h^0} & = &\left| \sigma_{ji}^{(2)} y_k^{(u)} \cos \theta_H \right|^2 \\
A_{H^0 H^0} & = &\left| \sigma_{ji}^{(3)} y_k^{(u)} \sin \theta_H \right|^2 \\
A_{h^0 H^0} & = &\text{ Re} \left[ \sigma_{ji}^{(2)}\sigma_{ji}^{(3)*}  
y_k^{(u)2} \sin \theta_H \cos \theta_H \right] \\
A_{A^0 A^0} & = &\left| \sigma_{ji}^{(4)} y_k^{(u)} \cos \beta \right|^2 \\
A_{G^0 G^0} & = &\left| \sigma_{ji}^{(5)} y_k^{(u)} \sin \beta \right|^2 \\
A_{A^0 G^0} & = &\text{ Re} \left[ \sigma_{ji}^{(4)}\sigma_{ji}^{(5)*}  
y_k^{(u)2} \sin \beta \cos \beta \right] \\
A_{Z h^0} & = &- \text{Re}\left[ \sigma_{ji}^{(2)} y_k^{(u)} \cos \theta_H
\left(\frac{8}{3} \sin^2 \theta_W - 1\right)\right] \\
A_{Z H^0} & = &- \text{Re}\left[ \sigma_{ji}^{(3)} y_k^{(u)} \sin \theta_H 
\left(\frac{8}{3} \sin^2 \theta_W - 1\right)\right] \\
A_{Z A^0} & = &- \text{Re}\left[ \sigma_{ji}^{(4)} y_k^{(u)} \cos \beta\right] \\
A_{Z G^0} & = &- \text{Re}\left[ \sigma_{ji}^{(5)} y_k^{(u)} \sin \beta\right] \ .
\end{eqnarray}

\subsection{Down-Type Quarks}

The decay width to down-type quarks is
\begin{eqnarray}
\lefteqn{\Gamma \left(\tilde{\ell}_i^- \to \tilde{\ell}_j^- d_k
\overline{d}_k\right) = \frac{9 m_i}{512 \pi^3} \left| \frac{g
\zeta_{ji}^{(2)}}{ \cos \theta_W} \right|^2 \sum_{t = 1}^2
B_{ZZ}^{(t)} I_{ZZ}^{(t)}} \nonumber \\
&& + \frac{9}{256 \pi^3 m_i}
\left( B_{h^0 h^0} I_{h^0 h^0}^r + B_{H^0 H^0} I_{H^0 H^0}^r + 2
B_{h^0 H^0} I_{h^0 H^0}^r \right. \nonumber \\
&& \left. \qquad + B_{A^0 A^0} I_{A^0 A^0}^p + B_{G^0 G^0} I_{G^0
G^0}^p + 2 B_{A^0 G^0} I_{A^0 G^0}^p \right) \nonumber \\
&& + 2 \left[ \frac{9 g \zeta_{ji}^{(2)}}{ 256 \sqrt{2} \pi^3 \cos
\theta_W} \left( B_{Z h^0} I_{Z h^0}^r + B_{Z H^0} I_{Z H^0}^r + B_{Z
A^0} I_{Z A^0}^p + B_{Z G^0} I_{Z G^0}^p \right) \right] \ ,
\end{eqnarray}
where again, the integrals are equivalent to those defined in
\appref{dileptonp} with quark masses substituted.  The coefficients
are
\begin{eqnarray}
B_{ZZ}^{(1)} & =& 
\frac{8}{9} \sin^4 \theta_W - \frac{4}{3} \sin^2 \theta_W + 1 \\
B_{ZZ}^{(2)} & =& \frac{8}{9} \sin^4 \theta_W - 
\frac{4}{3} \sin^2 \theta_W \\
B_{h^0 h^0} & = &\left| \sigma_{ji}^{(2)} y_k^{(d)} \sin \theta_H \right|^2 \\
B_{H^0 H^0} & = &\left| \sigma_{ji}^{(3)} y_k^{(d)} \cos \theta_H \right|^2 \\
B_{h^0 H^0} & = &- \text{Re} \left[ \sigma_{ji}^{(2)}\sigma_{ji}^{(3)*}  
y_k^{(d)2} \sin \theta_H \cos \theta_H \right] \\
B_{A^0 A^0} & = &\left| \sigma_{ji}^{(4)} y_k^{(d)} \sin \beta \right|^2 \\
B_{G^0 G^0} & = &\left| \sigma_{ji}^{(5)} y_k^{(d)} \cos \beta \right|^2 \\
B_{A^0 G^0} & = &- \text{Re} \left[ \sigma_{ji}^{(4)}\sigma_{ji}^{(5)*}  
y_k^{(d)2} \sin \beta \cos \beta \right] \\
B_{Z h^0} & = &- \text{Re}\left[ \sigma_{ji}^{(2)} y_k^{(d)} \sin \theta_H
\left(\frac{4}{3} \sin^2 \theta_W - 1\right)\right] \\
B_{Z H^0} & = &\text{ Re}\left[ \sigma_{ji}^{(3)} y_k^{(d)} \cos \theta_H 
\left(\frac{4}{3} \sin^2 \theta_W - 1\right)\right] \\
B_{Z A^0} & = &- \text{Re}\left[ \sigma_{ji}^{(4)} y_k^{(d)} \sin \beta\right] \\
B_{Z G^0} & = &\text{ Re}\left[ \sigma_{ji}^{(5)} y_k^{(d)} \cos \beta\right] \ .
\end{eqnarray}



\end{document}